\documentclass[12pt]{article}
\usepackage{latexsym,amsmath,amssymb,mathtools,multirow}
\usepackage{enumitem} 
\setlist{nolistsep}
\usepackage[square,comma,sort&compress,numbers]{natbib}
\usepackage{pstricks}  
\usepackage{graphicx}
\usepackage{hyperref}

\renewcommand{\a}{{\alpha}}

\newcommand{\dl}{{\delta}}
\newcommand{\eps}{{\epsilon}}

\newcommand{\m}{{\mu}}
\newcommand{\n}{{\nu}}

\newcommand{\pa}{{\partial}}

\newcommand{\sig}{{\sigma}}

\renewcommand{\th}{{\theta}}

\newcommand{\om}{{\omega}}
\newcommand{\Om}{{\Omega}}

\newcommand{\as}{\mathfrak{s}}

\newcommand{\ao}{\mathfrak{o}}

\newcommand{\rd}{\textnormal{d}}
\makeatletter
\renewcommand\section{\@startsection {section}{1}{\z@}%
    {-3.5ex \@plus -1ex \@minus -.2ex}%
    {2.3ex \@plus.2ex}%
    {\normalfont\large\bfseries}}
\newcommand{\vast}{\bBigg@{3}}
\newcommand{\Vast}{\bBigg@{8}}
\makeatother

\numberwithin{equation}{section}
\renewcommand{\theequation}{\arabic{section}.\arabic{equation}}

\begin{document}
\thispagestyle{empty}

\begin{center}
 
  {\huge Perturbative quantization of Yang-Mills theory} \\\vskip 6pt %
  {\huge with classical double as gauge algebra} %
  \vskip 45pt {F.~Ruiz~Ruiz} \vskip 3pt \emph{Departamento de F\'{\i}sica
    Te\'orica I, Universidad Complutense de Madrid\\ 28040 Madrid, Spain}

\end{center}

{\leftskip=30pt\rightskip=30pt
  
  \noindent Perturbative quantization of Yang-Mills theory with a gauge
  algebra given by the classical double of a semisimple Lie algebra is
  considered. The classical double of a real Lie algebra is a nonsemisimple
  real Lie algebra that admits a nonpositive definite invariant metric, the
  indefiniteness of the metric suggesting an apparent lack of unitarity.  It is
  shown that the theory is UV divergent at one loop and that there are no
  radiative corrections at higher loops. One-loop UV divergences are removed
  through renormalization of the coupling constant, thus introducing a
  renormalization scale. The terms in the classical action that would spoil
  unitarity are proved to be cohomologically trivial with respect to the
  Slavnov-Taylor operator that controls gauge invariance for the quantum
  theory. Hence they do not contribute gauge invariant radiative corrections
  to the quantum effective action and the theory is unitary. \\[15pt]%

  \noindent{\sc keywords:} Yang-Mills theory, classical double, BRS identity,
  Slavnov-Taylor operator, unitarity.

\par}

\vspace{30pt}
\section{Introduction}

Nonreductive metric Lie algebras are Lie algebras that (i) cannot be written
as a direct product of semisimple and Abelian Lie algebras but (ii) admit a
metric, where by a metric is meant a nondegenerate symmetric bilinear form
that is invariant under the adjoint action.  Here we will be concerned with
perturbative quantization of Yang-Mills theory for a particular class of such
algebras, known as classical doubles. These algebras describe the gauge
symmetries in a variety of problems, including three-dimensional
gravity~\cite{Achucarro-Townsend, Witten-three}, asymptotically flat solutions
to the Einstein equations in three and four dimensions~\cite{Barnich-Oblak-I,
  Barnich-Oblak-II, Barnich-Troessaert}, string actions in doubled
space~\cite{Hatsuda-Kamimura-Siegel}, or $1/N_{\textnormal{color}}$ expansions
for baryons in QCD~\cite{Dashen}.

Many WZW models~\cite{Nappi-Witten, Sfetsos-1, Sfetsos-2, Sfetsos-3,
  Mohammedi, FO-Stanciu-nonreductive, FO-Stanciu-double} based on nonreductive
Lie algebras, though of a different type, have a simpler structure than the
WZW models based on semisimple Lie algebras.  This suggests considering
Yang-Mills theories in four dimensions and investigate if the simplifications
introduced in two dimensions by going nonreductive carry through to four
dimensions. The problem was undertaken in Ref.~\cite{Tseytlin} for a class of
nonreductive algebras called double extensions. One-loop radiative corrections
for certain models were computed and it was argued that, if renormalizability
is assumed, there would not be higher-loop corrections.  An apparent lack of
unitarity was found.

The classical double, we denote it as $\mathfrak{g}_\ltimes$, of any real Lie
algebra $\mathfrak{g}$ is a Lie algebra of dimension twice the dimension of
$\mathfrak{g}$ that admits a metric. This metric determines a Yang-Mills
Lagrangian for the field that results from gauging the algebra. For
$\mathfrak{g}$~simple, the self-antiself dual instantons of the
$\mathfrak{g}_\ltimes$ Yang-Mills theory in four-dimensional Euclidean space
have been studied elsewhere~\cite{FRR}.  Every $\mathfrak{g}_\ltimes$
instanton has an embeded $\mathfrak{g}$ instanton with the same instanton
number and twice the number of collective coordinates. This doubling of
degrees of freedom and the simpler structure of classical doubles as compared
to double extensions suggest considering perturbative quantization of
Yang-Mills theory with gauge algebra $\mathfrak{g}_\ltimes$.

The theory is shown to have UV divergences at one loop but no radiative
corrections at higher loops.  As in the classical case, the physical degrees
of freedom of the quantum theory are doubled with respect to the
$\mathfrak{g}$ theory. In particular, the first and only coefficient (since
there are no radiative corrections beyond one loop) of the beta function is
twice that of the $\mathfrak{g}$ theory.  To disentangle truly gauge
invariant one-loop corrections from those due to gauge fixing, the
Slavnov-Taylor operator for the quantum theory is used.  The term in the
classical action that would spoil unitarity is cohomologically trivial with
respect to the Slavnov-Taylor operator, so in the quantum effective action it
can be put to zero through a field redefinition and poses no problem for
unitarity.

The manuscript is organized as follows. Section~2 contains a brief reminder of
classical doubles~$\mathfrak{g}_\ltimes$ and their Lie
groups~$\textnormal{G}_\ltimes$.  Classical Yang-Mills theory with gauge
group~$\textnormal{G}_\ltimes$ is formulated in Section 3, and the path
integral generating the theory's Green functions is derived. With the mind set
in avoiding miscounting the theory's degrees of freedom, special consideration
is given to gauge fixing, and three derivations for the gauge fixing terms in
Landau gauge are presented. Section 3 also discusses the emergence of the
classical theory as a limit of Yang-Mills theory with gauge algebra the direct
product $\mathfrak{g}\times\mathfrak{g}$. Section~4 contains our perturbative
analysis, with the calculation of one-loop 1PI radiative corrections and the
the proof that there are no higher-loop corrections. The one-loop divergences
are removed in Section~5 by adding a gauge invariant counterterm consistent
with unitarity, so the one-loop contribution to the quantum effective action
is positive definite. We conclude in Section~6.

\section{Semidirect products of Lie algebras and 
their groups} 

We start by reminding the definitions of classical double of a Lie algebra and
its Lie group.  Consider an arbitrary Lie algebra $\mathfrak{g}$ of dimension
$n$ with basis $\{T_a\}$ and commutation relations
$[T_a,T_b]=f_{ab}{}^c\,T_c$. As a vector space, $\mathfrak{g}$ has a dual
vector space $\mathfrak{g}^*$. Take on $\mathfrak{g}^*$ the canonical dual
basis $\{Z^a\}$, given by $Z^a(T_b)=\dl^a{}_b$.  The classical double of
$\mathfrak{g}$, denoted by $\mathfrak{g}_\ltimes$, is the Lie algebra of
dimension $2n$ with basis $\{T_a,Z^b\}$ and commutators
\begin{equation}
   [T_a,T_b]=f_{ab}{}^c\,T_c\,,\quad
   [T_a,Z^b]=-f_{ac}{}^b\,{Z}^c\,, \quad 
   [Z^a,Z^b]=0\,.
\label{SDalgebra}
\end{equation}
It is trivial to check that these commutators satisfy the Jacobi identity, so
indeed they define a Lie bracket. In a more mathematical
language~\cite{FO-Stanciu-double}, the algebra $\mathfrak{g}_{{\ltimes}}$ is
the semidirect product $\mathfrak{g}\ltimes\mathfrak{g}^*$ obtained by the
coadjoint action of $\mathfrak{g}$ on $\mathfrak{g}^*$.

It is clear from the commutators~(\ref{SDalgebra}) that
$\mathfrak{g}_{{\ltimes}}$ is not semisimple, so its Killing form is
degenerate and cannot be used as a metric. For convenience we recall that a
metric on a Lie algebra is a symmetric, nondegenerate, bilinear form $\Om$ such that
\begin{equation}
    \Om\,(A\,,[B,C])\,=\Om\,(\,[A,B]\,,C) 
\label{invariance}
\end{equation}
for all $A,\,B,\,C$ in the algebra. The relevance of
condition~(\ref{invariance}) is that it implies invariance under the adjoint
action of the algebra's Lie group,
\begin{equation}
  \Om\,(e^{-C}A\,e^{C},\,e^{-C}\/B\,e^{C}) = \Om\,(A,B)\,.
\label{adjointinvariance}
\end{equation} 
It is straightforward to check that, for $\{T_a,Z^b\}$ above, the bilinear
symmetric form
\begin{equation}
\begin{tabular}{cccccc}
             & &  & $T_b$ & $Z^b$ & \\[3pt]
\multirow{2}{*}{$\Om=$} 
          & $T_a$ & \multirow{2}{*}{$\Bigg(\!\!\!\!$} & $\om_{ab}$ 
                   &$\dl_a{}^b$ &\multirow{2}{*}{$\!\!\!\!\Bigg)$} \\[4.5pt]
          & $Z^a$ &   &$\dl_a{}^b$  & 0 & 
\end{tabular}
\label{SDmetric}
\end{equation}
is nondegenerate and solves equation~(\ref{invariance}), hence is a metric on
$\mathfrak{g}_{\ltimes}$. Here $\om_{ab}=\om(T_a,T_b)$ is an arbitrary
invariant, symmetric, bilinear form on $\mathfrak{g}$, that may be
degenerate. It is worth remarking that $\mathfrak{g}$ is arbitrary and need
not be metric itself.

Denote by ${\rm G}_{\ltimes},\,{\rm G}$ and ${\rm N}$ the Lie groups obtained
by exponentiation of the algebras $\mathfrak{g}_{\ltimes},\,\mathfrak{g}$ and
$\mathfrak{g}^*$. Every element $h$ of $\textnormal{G}$ and every $n$ of
$\textnormal{N}$ can be uniquely written as $h=e^T$ and $n=e^Z$, for some $T$
in $\mathfrak{g}$ and some $Z$ in $\mathfrak{g}^*$. In turn, every element $g$
of ${\rm G}_{\ltimes}$ can be uniquely written as $g=hn$. The product
$g_3=g_1g_2$ of two elements $g_1=h_1n_1$ and $g_2=h_2n_2$ of ${\rm
  G}_{\ltimes}$ is given by $g_3=h_3n_3$, with $h_3=h_1h_2$ and
$n_3=(h_2^{-1}n_1h_2)n_2$. It is clear that $g_3$ is in ${\rm G}_{\ltimes}$
since the Campbell-Hausdorff formula and the commutators~(\ref{SDalgebra})
imply that $h_3$ is in ${\rm G}$ and $n_3$ is in ${\rm N}$. It is very easy to
see that the group ${\rm G}_{\ltimes}$ is the semidirect product of ${\rm G}$
with the normal Abelian subgroup ${\rm N}$. Since ${\rm N}$ is isomorphic to
$\mathbf{R}^n$, the group ${\rm G}_{\ltimes}$ is noncompact.
 
So far no restriction has been placed on $\mathfrak{g}$. Assume now that it is
semisimple.  In this case,~$\mathfrak{g}_{\ltimes}$ can be viewed as a limit
of the direct product of $\mathfrak{g}$ with itself~\cite{FRR}. To see this,
take $\om_{ab}$ in eq.~(\ref{SDmetric}) proportional to the Killing form of
$\mathfrak{g}$. Since $\om_{ab}$ is nondegenerate, indices in the structure
constants $f_{ab}{}^c$ can be raised and lowered using $\om_{ab}$ and its
inverse $\om^{ab}$, defined by $\om^{ab}\om_{bc}=\dl^a{}_c$. This gives
completely antisymmetric structure constants
\begin{equation*}
     f_{abc\!} =f_{ab}{}^d\,\om_{dc}\,,\qquad f_{abc}=-f_{bac}=f_{bca}\,.
\end{equation*}
Perform in $\mathfrak{g}^*$ the change of generators\, $Z^a\to
Z_a=\om_{ab}Z^b$. In the basis $\{T_a,Z_b\}$ the Lie bracket~(\ref{SDalgebra})
reads
\begin{equation}
      [T_a,T_b]=f_{ab}{}^c\,T_c\,,\quad
      [T_a,Z_b]=f_{ab}{}^c\,{Z}_c\,, \quad 
      [Z_a,Z_b]=0
\label{undeformedSDalgebra}
\end{equation}
and the metric $\Om$ is recast as
\begin{equation}
\begin{tabular}{cccccc}
             & &  & $T_b$ & $Z_b$ & \\[3pt]
\multirow{2}{*}{$\Om=$} 
          & $T_a$ & \multirow{2}{*}{$\Bigg(\!\!\!\!$} & $\om_{ab}$ 
                   & $\om_{ab}$ &\multirow{2}{*}{$\!\!\!\!\Bigg)$\,.} \\[4.5pt]
          & $Z_a$ &   &$\om_{ab}$  & 0 & 
\end{tabular}
\label{undeformedSDmetric}
\end{equation}
Consider now the commutators
\begin{equation}
      [T_a,T_b]=f_{ab}{}^c\,T_c\,,\quad
      [T_a,Z_b]=f_{ab}{}^c\,{Z}_c\,, \quad 
      [Z_a,Z_b]=t^2 f_{ab}{}^c\, T_c\,,
\label{deformedSDalgebra}
\end{equation}
where $t$ is an arbitrary real parameter.  It is trivial to show that
they satisfy the Jacobi identity for all $t$. Hence they define a Lie algebra,
call it $\mathfrak{g}_{\ltimes}^{\,t}$, that reduces to
$\mathfrak{g}_{\ltimes}$ in the limit $t\to 0$.  Furthermore,
$\mathfrak{g}_{\ltimes}^{\,t}$ admits the invariant metric
\begin{equation}
\begin{tabular}{cccccc}
             & &  & $T_b$ & $Z_b$ & \\[3pt]
\multirow{2}{*}{$\Om^{\,t}=$} 
          & $T_a$ & \multirow{2}{*}{$\Bigg(\!\!\!\!$} & $\om_{ab}$ 
                   & $\om_{ab}$ &\multirow{2}{*}{$\!\!\!\!\Bigg)$\,.} \\[4.5pt]
          & $Z_a$ &   &$\om_{ab}$  & $t^2 \om_{ab}$ & 
\end{tabular}
\label{deformedSDmetric}
\end{equation}
The change of basis
\begin{equation}
       X^\pm_a= \frac{1}{2}\,\Big( T_a\pm \frac{1}{t}\,Z_a\Big)\,,\quad
\end{equation}
in $\mathfrak{g}_{\ltimes}^{\,t}$ 
transforms the Lie bracket~(\ref{deformedSDalgebra}) in
\begin{equation}
      [X^\pm_a,X^\pm_b]=f_{ab}{}^c\,X^\pm_c\,,\quad
      [X^+_a,X^-_b]=0
\label{deformedSDalgebraXY}
\end{equation}
and
gives for the metric $\Om^t$ in eq.(\ref{deformedSDmetric}) the block diagonal form
\begin{equation}
\begin{tabular}{cccccc}
             & &  & $X^+_b$ & $X^-_b$ & \\[3pt]
   \multirow{2}{*}{$\Om^{\,t}=$} 
         & $X^+_a$ & \multirow{2}{*}{$\vast(\!\!\!\!$} 
          & $\frac{1}{2}\,\big(1+\frac{1}{t}\big)\,\om_{ab}$ 
                  & 0 &\multirow{2}{*}{$\!\!\!\!\vast)$.} \\[4.5pt]
          & $X^-_a$ & & 0 &  $\frac{1}{2}\,\big(1-\frac{1}{t}\big)\,\om_{ab}$ & \\[3pt]
\end{tabular}
\label{deformedSDmetricXY}
\end{equation}
It is clear from eqs.~(\ref{deformedSDalgebraXY})
and~(\ref{deformedSDmetricXY}) that $\mathfrak{g}_{\ltimes}^{\,t}$ is the
direct product $\mathfrak{g}\times\mathfrak{g}$, so its Lie group ${\rm G
}_{\ltimes}^{\,t}$ is the direct product ${\rm G }\times{\rm G}$.

\section{The gauge fixed classical action}

We assume from now on that $\mathfrak{g}$ is semisimple and work in the basis
$\{T_a,Z_b\}$ of $\mathfrak{g}_{\ltimes}$.  The commutation relations are as
in eq.~(\ref{undeformedSDalgebra}) and the metric $\Om$ as
in~(\ref{undeformedSDmetric}). The Lie groups of $\mathfrak{g}$ and
$\mathfrak{g}_{\ltimes}$ will be denoted by $\textnormal{G}$ and
$\textnormal{G}_{\ltimes}$. All quantities taking values in
$\mathfrak{g}_{\ltimes}$ will be written in boldface, whereas their $T_a$ and
$Z_a$ components will be labeled with subscripts~{\sc t} and~{\sc z}. For the
gauge field $\boldsymbol{A}_\m$ and its field strength\,
$\boldsymbol{F}_{\m\n} =\pa_\m \boldsymbol{A}_{\n} - \pa_\n
\boldsymbol{A}_{\m} + [\boldsymbol{A}_\m ,\boldsymbol{A}_\n ]$, we have
\begin{align*}
    \boldsymbol{A}_\m & = A^a_{\textnormal{\sc T}\m} T_a 
         + A^a_{\textnormal{\sc Z}\m} Z_a\,,    \\[3pt] 
     \boldsymbol{F}_{\m\n} &=  F^a_{\textnormal{\sc T}\m\n} T_a 
     + F^a_{\textnormal{\sc Z}\m\n} Z_a  \,. 
\end{align*}
The expressions of $F^a_{\textnormal{\sc T}\m\n}$ and $F^a_{\textnormal{\sc
    Z}\m\n}$ in terms of $A^a_{\textnormal{\sc T}\m}$ and
$A^a_{\textnormal{\sc Z}\m}$ follow from the
commutators~(\ref{undeformedSDalgebra}),
\begin{align*}
    F^a_{\textnormal{\sc T}\m\n} & = \pa_\m A^a_{\textnormal{\sc t}\n} 
           -  \pa_\n A^a_{\textnormal{\sc T}\m}  
          + f_{bc}{\!}^a\, A^b_{\textnormal{\sc T}\m}\,          
             A^c_{\textnormal{\sc Z}\n}\,,    \\[3pt]
    F^a_{\textnormal{\sc Z}\m\n} & = \pa_\m A^a_{\textnormal{\sc Z}\n} 
        -  \pa_\n A^a_{\textnormal{\sc Z}\m}
        + f_{bc}{\!}^a \,\big( 
                A^b_{\textnormal{\sc T}\m}\, A^a_{\textnormal{\sc Z}\n}  
             - A^b_{\textnormal{\sc T}\n}\, A^a_{\textnormal{\sc Z}\n}
             \big)\,.    \label{FZ}  
\end{align*}
The action of the covariant derivative\, $\boldsymbol{D}_\m{\!} = \pa_\m +
[\boldsymbol{A}_\m,~]$ on any tensor $\boldsymbol{\Phi}=
\Phi^a_{\textnormal{\sc t}} T_a +\Phi^a_{\textnormal{\sc z}} Z_a$ has
components
\begin{align*}
   \big(\boldsymbol{D}_\m\boldsymbol{\Phi}\big)^a_{\textnormal{\sc T}}& 
       =  \pa_\m\, \Phi^a_{\textnormal{\sc T}} 
       + f_{bc}{\!}^a\,A^b_{\textnormal{\sc T}\m}\,\Phi^c_{\textnormal{\sc T}}
       \,,   \\[3pt] 
   (\boldsymbol{D}_\m\boldsymbol{\Phi})^a_{\textnormal{\sc Z}}& 
       =  \pa_\m \Phi^a_{\textnormal{\sc Z}} + f_{bc}{\hspace{-0.5pt}}^a\,
                  A^b_{\textnormal{\sc T}\m} \,\Phi^c_{\textnormal{\sc Z}}  
       + f_{bc}{\hspace{-0.5pt}}^a\,A^b_{\textnormal{\sc Z}\m} \, 
                 \Phi^c_{\textnormal{\sc T}} \,.
\end{align*}
On the right hand side of the first equation, one recognizes the covariant
derivative for the algebra~$\mathfrak{g}$, which we denote by~$D_\m$,
\begin{equation}
    D_{c}{\hspace{-0.5pt}}^a{\hspace{0.5pt}}_\m := \dl_c{\!}^a\,\pa_\m 
              + f_{b\hspace{0.5pt}c}{\!}^a\,A^b_{\textnormal{\sc T}\m}\,.
\label{cov-der}
\end{equation}
Finite gauge transformations read
\begin{equation}
   \begin{array}{rcl}
   \boldsymbol{A}_\m & \to & \boldsymbol{A}^\prime_\m = g^{-1}\,\pa_\m\/g 
              + g^{-1}\boldsymbol{A}_\m\,g\,,    \\[3pt]
  \boldsymbol{F}_{\m\n}  & \to & \boldsymbol{F}^{\prime}_{\m\n} 
        = g^{-1} \boldsymbol{F}_{\m\n}\, g\,, 
  \end{array}
\label{GT-F}
\end{equation}
with $g$ an arbitrary group element in ${\rm G}_{\ltimes}$,
\begin{equation*}
   g(x)=\exp\!\big[ \th^a_{\textnormal{\sc T}}(x)\hspace{0.5pt}T_a\big] \>
        \exp\!\big[ \th^a_{\textnormal{\sc Z}}(x)\hspace{0.5pt}Z_a \big]\,.
\end{equation*}
Infinitesimally these transformations take the form
\begin{align}
   \dl A^a_{\textnormal{\sc T}\m} & = D_\m\th^a_{\textnormal{\sc T}} \,,
    \label{GT-AT} \\[3pt]
   \dl A^a_{\textnormal{\sc Z}\m} & = D_\m\th^a_{\textnormal{\sc Z}} 
          + f_{bc}{\hspace{-0.5pt}}^a\,A^b_{\textnormal{\sc Z}\m}
              \th^c_{\textnormal{\sc T}} 
   \label{GT-AZ} 
\end{align}
for the gauge field, and
\begin{align}
   \dl F^a_{\textnormal{\sc T}\m\n} & = f_{bc}{\hspace{-0.5pt}}^a\, 
          F^b_{\textnormal{\sc T}\m\n} \th^c_{\textnormal{\sc T}}\,,
    \label{GT-FT} \\[3pt]
   \dl F^a_{\textnormal{\sc Z}\m\n} & =  f_{bc}{\hspace{-0.5pt}}^a\, 
     \big(  F^b_{\textnormal{\sc z}\m\n} \th^c_{\textnormal{\sc T}}
      + F^b_{\textnormal{\sc T}\m\n} \th^c_{\textnormal{\sc Z}} \big) 
   \label{GT-FZ} 
\end{align}
for the field strength. The invariance condition~(\ref{adjointinvariance}) for
the metric $\Om$ and the transformation law~(\ref{GT-F}) for the field
strength $\boldsymbol{F}_{\m\n}$ imply that the Lagrangian density
\begin{equation}
   {\cal L}_{\textnormal{\sc ym}} = \frac{1}{4g^2}~
          \Om \big(\boldsymbol{F}_{\m\n},\boldsymbol{F}^{\m\n}\big)
       =  \frac{1}{4g^2}~\om_{ab}\, \big(\, 
           F^a_{\textnormal{\sc T}\m\n} F^{b\m\n}_{\textnormal{\sc T}}
       + 2\, F^a_{\textnormal{\sc T}\m\n}\, F^{b\m\n}_{\textnormal{\sc Z}}\,\big)\,,
\label{YM-L}
\end{equation}
where $g$ is a coupling constant, is gauge invariant. We are interested in
perturbatively quantizing Yang-Mills theory with Lagrangian ${\cal
  L}_{\textnormal{\sc ym}}$. To obtain a path integral that generates the
theory's Green functions, we next fix the gauge. We do this in three different
ways.

\medskip {\bf Gauge fixing I}. Introduce a ghost field $\boldsymbol{c}$, an
antighost field $\boldsymbol{\bar{c}}$ and a Lagrange multiplier field
$\boldsymbol{b}$,
\begin{align*}
   \boldsymbol{c} & =   c^a_{\textnormal{\sc T}}\, T_a
          + c^a_{\textnormal{\sc Z}}\,Z_a \,,  \\[3pt] 
   \boldsymbol{\bar{c}}  & = \bar{c}^a_{\textnormal{\sc T}}\,T_a
          + \bar{c}^a_{\textnormal{\sc Z}} \,Z_a \,, \\[3pt] 
 \boldsymbol{b}  & = b^a_{\textnormal{\sc T}} \,T_a
          + b^a_{\textnormal{\sc Z}} \,Z_a  \,. 
\end{align*}
Use the infinitesimal form of the gauge transformations to define a BRS
operator $s$ by its action on the fields,
\begin{equation}
   s\boldsymbol{A}_\m =\boldsymbol{D}_\m \boldsymbol{c}\,, \quad 
   s\boldsymbol{c}= - \boldsymbol{cc}\,,    \quad 
   s\boldsymbol{\bar{c}}= \boldsymbol{b}\,,  \quad 
   s\boldsymbol{b}= 0.\,
\label{BRS-op}
\end{equation}
The operator $s$ commutes with $\pa_\m,\,\boldsymbol{A}_\m$ and
$\boldsymbol{b}$, and anticommutes with $\boldsymbol{c}$ and
$\boldsymbol{\bar{c}}$.  The BRS transformations for the {\sc t} and {\sc z} components of
the fields can be obtained either from the definition of $s$ in
eq.~(\ref{BRS-op}) and the commutations relations~(\ref{undeformedSDalgebra}),
or directly from eqs.~(\ref{GT-AT}) and~(\ref{GT-AZ}). They read
\begin{equation}
       sA_{\textnormal{\sc T}\m} 
            = D_\m\hspace{0.5pt} c^a_{\textnormal{\sc T}}\,,\quad 
       s c^a_{\textnormal{\sc T}} = -\, \frac{1}{2}\> f_{bc}{}^a\, 
             c^b_{\textnormal{\sc T}}\, c^c_{\textnormal{\sc T}}\,,\quad
      s \bar{c}^a_{\textnormal{\sc T}} = b^a_{\textnormal{\sc T}}\,,\quad
      sb^a_{\textnormal{\sc T}} = 0\,,
\label{BRS-T}
\end{equation}
and 
\begin{equation}
     s A^a_{\textnormal{\sc Z}\m} = D_\m\hspace{0.5pt}  c^a_{\textnormal{\sc Z}} 
              +  f_{bc}{}^a\,A^b_{\textnormal{\sc Z}\m} c^c_{\textnormal{\sc T}} \,,
      \quad
    s  c^a_{\textnormal{\sc Z}} = - f_{bc}{}^a \, c^b_{\textnormal{\sc T}} 
             c^c_{\textnormal{\sc Z}} \,, \quad
    s  \bar{c}^a_{\textnormal{\sc Z}} = b^a_{\textnormal{\sc Z}} \,, \quad
   sb^a_{\textnormal{\sc Z}} =0\,.
\label{BRS-Z}
\end{equation}
Using eqs.~(\ref{BRS-op}), or their equivalent eqs.~(\ref{BRS-T}) and~(\ref{BRS-Z}),
it is very easy to check that $s^2\!=0$.  Note that the BRS
transformations~(\ref{BRS-T}) are the same as for the semisimple gauge algebra
$\mathfrak{g}$. In Lorenz gauge, the gauge fixing Lagrangian is the BRS
variation
\begin{equation*}
   {\cal L}_{\textnormal{\sc gf}} =  - \frac{1}{g^2}\> s\,\Om\hspace{0.5pt} 
     \Big(\hspace{0.5pt} \boldsymbol{\bar{c}} \,,\,  
     \frac{\a}{2}\>\boldsymbol{b} + \pa\boldsymbol{A}\Big)\,,
\end{equation*}
where $\a$ is the gauge parameter and a contraction of the spacetime indices
in $\pa_\m$ and $\boldsymbol{A}_\m$ is understood. Expanding in terms of 
field components, ${\cal L}_{\textnormal{\sc gf}} $ becomes
\begin{align}
     {\cal L}_{\textnormal{\sc gf}} =  \>\frac{1}{g^2}\> \om_{ab} \,\Big[ & - \,
         \frac{\a}{2}\> ( \,b^a_{\textnormal{\sc T}}\,b^b_{\textnormal{\sc T}}
                  + 2\ b^a_{\textnormal{\sc T}}\,b^b_{\textnormal{\sc Z}}\,)       
        - b^a_{\textnormal{\sc T}}\,( \pa A^b_{\textnormal{\sc T}}\,) 
        - b^a_{\textnormal{\sc T}}\,( \pa A^b_{\textnormal{\sc Z}}) 
        - b^a_{\textnormal{\sc Z}}\,( \pa A^b_{\textnormal{\sc T}})
        \nonumber \\[6pt]
    & + \bar{c}^a_{\textnormal{\sc T}}\,
           \big( \pa D c^b_{\textnormal{\sc T}}\big)
        + \bar{c}^a_{\textnormal{\sc T}}\,
           \big( \pa D c^b_{\textnormal{\sc Z}}\big)
        + \bar{c}^a_{\textnormal{\sc T}}\, \pa\, 
           ( f_{cd}{}^b\,A^c_{\textnormal{\sc Z}}\,c^d_{\textnormal{\sc T}} )
        + \bar{c}^a_{\textnormal{\sc Z}}\,\big( \pa D c^b_{\textnormal{\sc T}}\big)\, \Big]\,,
\label{GF-L-comp}
\end{align}
with $D_\m$ the $\mathfrak{g}$ covariant derivative derivative in
eq.~(\ref{cov-der}).  The gauge fixed Lagrangian is the sum
\begin{equation}
  {\cal L}_{{\ltimes}} = {\cal L}_{\textnormal{\sc ym}}   
          + {\cal L}_{\textnormal{\sc gf}}\,.
\label{SD-L}
\end{equation}

Introduce external sources $\{\boldsymbol{J_\Phi}\}=\{\boldsymbol{J}^{\m\!},
\boldsymbol{\bar{\zeta}}, \boldsymbol{\zeta}, \boldsymbol{B}\}$ \,for the
fields $\{\boldsymbol{\Phi}\}= \{\boldsymbol{A}_\m, \boldsymbol{c},
\boldsymbol{\bar{c}}, \boldsymbol{b}\}$, with
\begin{equation*}
  \boldsymbol{J_\Phi} =  J_\Phi{}^a_{\textnormal{\sc T}} \,T_a
      +  J_\Phi{}^a_{\textnormal{\sc Z}} \, Z_a\,.
\end{equation*}
The path integral that generates the theory's Green functions is given by
\begin{equation}
     Z\big[\boldsymbol{J}, \boldsymbol{\bar{\zeta}},  
              \boldsymbol{\zeta}, \boldsymbol{B}\big] 
     = \int [\rd \boldsymbol{A}]\> [\rd \boldsymbol{c}]\>
             [\rd \boldsymbol{\bar{c}}]\> [\rd \boldsymbol{b}]~
        \exp \hspace{-0.5pt}\Big[ \!-\! \int \!\rd^4\hspace{-0.5pt}x \>
                      {\cal L}_{{\ltimes}}  + S_{\rm ext} \Big]  \,,
\label{ZJ}
\end{equation}
where\, $[\rd\boldsymbol{\Phi}]=[\rd \Phi_{\textnormal{\sc T}}]\,
[\rd\Phi_{\textnormal{\sc Z}}]$ for every field $\boldsymbol{\Phi}$, and
$S_{\rm ext}$ is the source term
\begin{equation}
    S_{\rm ext} =  \frac{1}{g^2} \int \rd^4\hspace{-0.5pt}x\> \big[\, 
             \Om(\boldsymbol{J}, \boldsymbol{A}) 
         + \Om(\boldsymbol{\bar{\zeta}},\boldsymbol{c})  
         + \Om(\boldsymbol{\bar{c}},\boldsymbol{\zeta})    
         + \Om(\boldsymbol{B},\boldsymbol{b})\,  \big] \,.
\label{external}
\end{equation}
The external sources are coupled to fields through the Lie algebra metric
$\Om$, so that
\begin{equation*}
    \Om(\boldsymbol{J_\Phi}, \boldsymbol{\Phi}) 
       = \om_{ab}\,\big(\,
             J_\Phi{}^a_{\textnormal{\sc T}} \, \Phi^{b} _{\textnormal{\sc T}} 
         + J_\Phi{}^a_{\textnormal{\sc T}} \,\Phi^{b} _{\textnormal{\sc Z}} 
         + J_\Phi{}^a_{\textnormal{\sc Z}}\,\Phi^{b} _{\textnormal{\sc T}} \big) 
\end{equation*}
for every field $\boldsymbol{\Phi}$ and its source $\boldsymbol{J_\Phi}$. The
Green functions are obtained by functionally differentiating with respect to
the external sources. For example, $\langle A^a_{\textnormal{\sc T}\m}(x)
\hspace{0.5pt} A^b_{\textnormal{\sc Z}\n}(y) \rangle$ is given by
\begin{equation*}
  { \om^{ac}~ \frac{\dl}{{\dl {J^{}}_\textnormal{\sc Z}^{c\m}(x)}}~
    \om^{bd} \bigg[\, \frac{\dl}
                 {\dl {J^{}}_\textnormal{\sc T}^{d\hspace{0.75pt}\n}(y)} 
        - \frac{\dl}{\dl {J^{}}_\textnormal{\sc Z}^{d\hspace{0.75pt}\n}(y)} 
              \,\bigg]\> \ln\hspace{-.5pt} Z[\boldsymbol{J_\phi}]~ 
     \bigg\vert }_{\boldsymbol{J_\phi}=0}\,.
\end{equation*}

\medskip {\bf Gauge fixing II}. Equivalently, the path integral~(\ref{ZJ}) can
be derived as follows. In the naive path integral
\begin{equation}
  \int [\rd \boldsymbol{A}]~ \exp\Big[ -\!\int\! 
      \rd^4\hspace{-0.5pt}x~  {\cal L}_{\textnormal{\sc ym}} \Big]\,,
\label{naive}
\end{equation}
to avoid integrating over gauge equivalent degrees of freedom, replace 
\begin{equation}
     \int [\rd \boldsymbol{A}]~\to~ \int [\rd \boldsymbol{A}]~
      \dl\big(\pa\boldsymbol{A}- \boldsymbol{f}\big)~
    \Delta_{\pa\boldsymbol{A}}\,,
\label{measure}
\end{equation}
where the Dirac delta imposes the Lorenz gauge fixing condition 
$\pa\boldsymbol{A}=\boldsymbol{f}$ and
\begin{equation*} 
  \Delta_{\pa\boldsymbol{A}} = \textnormal{det} { \left[ 
    \frac{\dl}{\dl\boldsymbol{\th}(y)}~
       \pa\big( \boldsymbol{A} 
               + \dl\hspace{-1pt}\boldsymbol{A}\big)(x)
           \right]}_{\pa\boldsymbol{A}=0}
\end{equation*}
is the corresponding Faddeev-Popov determinant.  Proceed now as usual:
\begin{itemize}
\item[(i)] Average over~$\boldsymbol{f}$ with Gaussian type weight. That is,
  introduce in the measure
\begin{equation*}
   \int [\rd \boldsymbol{f}] \exp\Big[ -  
       \frac{1}{2\a g^2} \int\!\rd^4\hspace{-0.5pt}x\;
          \Om(\boldsymbol{f},\boldsymbol{f}) \,\Big]\,.
\end{equation*}
\item[ii)] Exponentiate $\dl\big(\pa\boldsymbol{A}- \boldsymbol{f}\big)$ by
  means of an auxiliary field $\boldsymbol{b}$, 
\begin{equation*}
   \dl\big(\pa\boldsymbol{A}- \boldsymbol{f}\big) = \int   
      [\rd \boldsymbol{b}]~ \exp\Big[ \frac{i}{g^2} \!\int\!
     \Om(\boldsymbol{b}, \pa\boldsymbol{A}-\boldsymbol{f})\,\Big] \,.
\end{equation*}
\item[(iii)] Write the determinant 
\begin{equation*} 
    \Delta_{\pa\boldsymbol{A}} = 
      \textnormal{det}~\dl^{(4)}(x-y) \begin{pmatrix}
     \pa^{\hspace{.5pt}\m} D_c{\hspace{-1pt}}^a{\hspace{-1.5pt}}_\m (x) & 0 \\[3pt]
  f_{bc}{\hspace{-0.5pt}}{}^a\,\pa^{\hspace{.5pt}\m} 
          A_{\textnormal{\sc Z}\m}^b(x) & \pa^{\hspace{.5pt}\m}
        D_c{\hspace{-1pt}}^a{\hspace{-1.5pt}}_\m(x) \end{pmatrix}
\end{equation*}
as a path integral over Grassmann fields $c_1^a,\,c_2^a$ and
$\bar{c}_{1a},\,\bar{c}_{2a}$,
\begin{equation*}
   \Delta_{\pa\boldsymbol{A}\!}  =  \int [\rd \bar{c}_1]\,  [\rd\bar{c}_2]\,  
    [\rd c_1]\,  [\rd c_2] ~
     \exp\! \Big[ - \>\frac{1}{g^2}\! \int\! 
      \rd^4\hspace{-0.5pt}x \>   \big(\, \bar{c}_{1a}\,\pa D c_1^a 
          + \bar{c}_{2a} \,f^a{\!}_{cb}\,\pa_\m 
                      (A^{c\m}_{\textnormal{\sc T}} c^b_1) 
          + \bar{c}_{2a}\,\pa D c_2^a \big) \Big] \,.
\end{equation*}
\end{itemize}
Use next (i)-(iii) in eq.~(\ref{measure}), integrate over $\boldsymbol{f}$, make
the change\footnote{A more precise notation for the {\sc t} and {\sc z}
  components of a $\mathfrak{g}_\ltimes$-valued field is $\boldsymbol{\Phi} =
  \Phi^{T_a}T_a + \Phi^{Z_a}Z_a$. Not to load the writing, we have used
  instead $\Phi^a_{\textnormal{\sc T}\!}:=\Phi^{T_a}$ and
  $\Phi^a_{\textnormal{\sc Z}\!}:=\Phi^{Z_a\!}$. Using $\Om$ and $\Om^{-1}$ to
  lower and raise indices, one has
\begin{alignat*}{4}
     \Phi_{T_a} & = \om_{ab}\,( \Phi^b_{\textnormal{\sc T}}  
                              +  \Phi^b_{\textnormal{\sc Z}}) \,,& \qquad 
     \Phi_{Z_a} & =  \om_{ab}\,\Phi^b_{\textnormal{\sc T}} \,\\[3pt]
     \Phi^a_{\textnormal{\sc T}} &  = \om^{ab}\, \phi_{Z_b}  \,, &  
     \Phi^a_{\textnormal{\sc Z}} & 
                  = \om^{ab}\,(\Phi_{T_b} - \Phi_{Z_b}) \,.
\end{alignat*}
Hence $(\bar{c}_{1a} , \bar{c}_{2a})$ in eq.~(\ref{identifications}) is
nothing but $(\bar{c}_{T_a}, \bar{c}_{Z_a})$.}
\begin{alignat}{4}
     c^a_{\textnormal{\sc T}} & = c_1^a\,, &\qquad 
    c^a_{\textnormal{\sc Z}} & =c_2^a\,\nonumber \\[3pt]
     \bar{c}^a_{\textnormal{\sc T}} & =\om^{ab}\,\bar{c}_{2b}\,, &
     \bar{c}^a_{\textnormal{\sc Z}} & 
                =\om^{ab}\,(\, \bar{c}_{1b} - \bar{c}_{2b})\,,
\label{identifications}
\end{alignat}
and replace~$\boldsymbol{b}$ with~$i\boldsymbol{b}$. This gives for the path
integral the expression in eq.~(\ref{ZJ}).

\medskip {\bf Gauge fixing III}. The observation at the end of Section~2
concerning deformations of~$\mathfrak{g}_{\ltimes}$ suggests that classical
${\rm G}_{\ltimes}$ Yang-Mills theory can be regarded as a limit of ${\rm
  G}\times{\rm G}$ Yang-Mills theory.  To see this, consider two copies of a
Yang-Mills theory, both with gauge group the semisimple Lie group~${\rm
  G}$. Label the copies with the subscripts~$+$ and~$-$, so that
$A^a_{\pm\hspace{1pt}\mu},\, F^a_{\pm\hspace{1pt}\mu\nu}$ and $g_\pm$ denote
their gauge fields, field strengths and coupling constants.  Fix in both
copies a Lorenz gauge by introducing ghost, antighost and auxiliary fields
$c^a_\pm,\,\bar{c}^a_\pm$ and $b^a_\pm$, and a BRS operators $s$ given by
\begin{equation*}
       s A^a_{\pm\hspace{0.5pt}\m} 
           = D_{\pm\hspace{0.5pt}\mu}\hspace{0.5pt} c^a_\pm \,,\quad 
       s c^a_\pm = -\, \frac{1}{2}\> f_{bc}{}^a\, c^b_\pm \, c^c_\pm \,,\quad
      s \bar{c}^a_\pm = b^a_\pm \,,\quad
      s b^a_\pm = 0\,.
\end{equation*}
The corresponding gauge fixed Lagrangians read
\begin{equation*}
   {\cal L}_\pm = \frac{1}{4 g^2_\pm} ~ \om_{ab}\, 
                     F^{\,a}_{\pm\, \m\n}\, F^{\,b\,\m\n\!}_{\,\pm}
       -  \, \frac{1}{g^2_\pm} ~\om_{ab} \, s\hspace{0.5pt}
                  \Big[ \hspace{0.5pt}  \bar{c}^{\,a}_\pm\hspace{0.5pt} \Big( 
                \frac{\a_\pm}{2}~b^{\,b}_\pm + \pa A^{\,b}_{\pm}\, \Big)  \Big] \,.
\end{equation*}
The sum
\begin{equation*}
      {\cal L}_{+-} = {\cal L}_+ + {\cal L}_- 
\end{equation*}
describes two theories not interacting with each other. Write $g_\pm$ in terms
of a coupling constant $g$ and a parameter $t$ as
\begin{equation*}
  \frac{1}{g^2_\pm} = \frac{1}{2g^2} ~\Big( 1\pm\frac{1}{t}\Big)\,,
\end{equation*}
and introduce fields $\Phi^a_{\textnormal{\sc T}}$ and
$\Phi^a_{\textnormal{\sc Z}}$ given by
\begin{equation*}
  \Phi^a_{\textnormal{\sc T}} =\frac{1}{2}\>(\Phi^a_+ + \Phi^a_-)\,,\quad
   \Phi^a_{\textnormal{\sc Z}} = \frac{1}{2t}\>(\Phi^a_+ - \Phi^a_-)
\end{equation*}
for $\Phi_\pm^a=A^a_{\pm\hspace{1pt}\m},\, c^{a\hspace{-.5pt}}_\pm,\,
\bar{c}^{a\hspace{-.5pt}}_\pm,\,b^a_\pm$, Take now $\a_+\!=\a_-\!=\a$ and send
$t\to 0$.  In this limit, the~$+$ and~$-$ sectors couple and the Lagrangian
${\cal L}_{+-}$ becomes the Lagrangian ${\cal L}_\ltimes$ of the ${\rm
  G}_\ltimes$ theory given by eqs.~(\ref{YM-L}),~(\ref{GF-L-comp})
and~(\ref{SD-L}).

\section{Radiative corrections}

To explicitly calculate radiative corrections, we take in this section
$\mathfrak{g}$ to be $\mathfrak{su}\textnormal{(N)}$. The group ${\rm G}$ is
then $\textnormal{SU(N)}$ and for ${\rm G}_{\ltimes}$ we use the notation
$\textnormal{SU(N)}_{\ltimes}$.

In the conventions of Section 2, in which a group element $h$ of
$\textnormal{SU(N)}$ is written as $e^T$, the elements $T$ of
$\mathfrak{su}(N)$ in the defining (fundamental) representation are traceless
antihermitean matrices.  We normalize the structure constants $f_{ab}{}^c$ of
$\mathfrak{su}(N)$ by requiring\,
$f_{ca}{\hspace{-.5pt}}^df_{db}{\hspace{-.5pt}}^c={\rm N}\dl_{ab}$. This gives
for the Killing form $k_{ab}={\rm N}\dl_{ab}$ and amounts to taking\, ${\rm
  tr}\hspace{0.5pt}[\hspace{0.5pt}T_{(R)a}
T_{(R)b}\hspace{0.5pt}]=C_2\hspace{0.5pt}\dl_{ab}$ in a representation $R$,
with $C_2={\rm N}$ in the adjoint representation and $C_2=-1/2$ in the
defining representation. For $\om_{ab}$ in eq.~(\ref{undeformedSDmetric}), we
take
\begin{equation}
  \om_{ab}=  \dl_{ab}\,,
\label{Killing}
\end{equation}
so indices in $f_{ab}{\hspace{-0.5pt}}^c$ are lowered and raised with
$\dl_{ab}$ and $\dl^{ab}$.
\begin{figure}[ht]
\begin{center}
\rule{0.85\textwidth}{0.4pt}\\[6pt]
\includegraphics[scale=0.8]{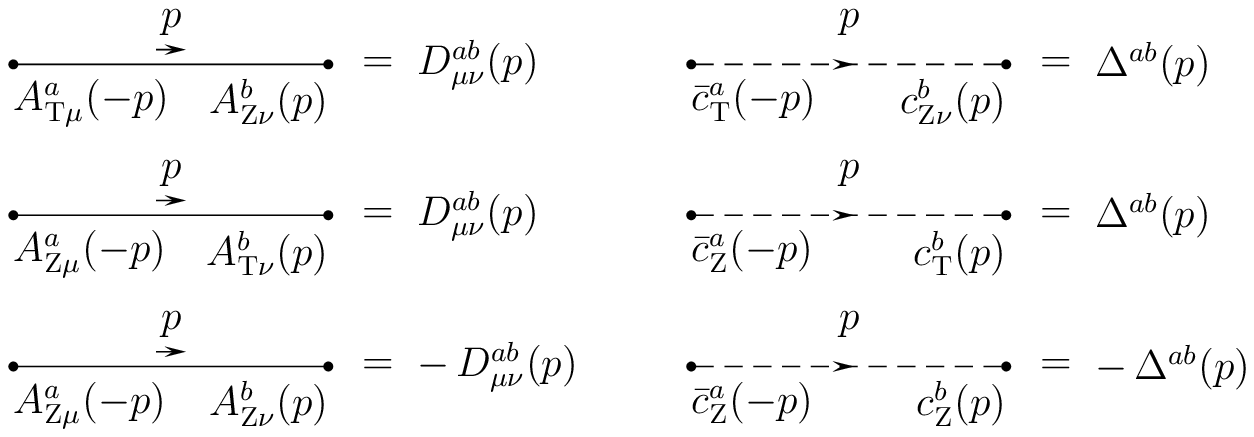}  
\caption{\it Free propagators for\, $\textnormal{SU(N)}_{\displaystyle{\ltimes}\!}$
 Yang-Mills in Lorenz gauge.}
\rule{0.85\textwidth}{0.4pt}
\end{center}
\vspace{-12pt}
\label{fig:propagators}
\end{figure}

\medskip {\bf Feynman rules in Lorenz gauge}. Introduce external sources
$\boldsymbol{K}^{\m\!}$ and $\boldsymbol{H}$ for the nonlinear BRS transforms
$s\hspace{-1.5pt}\boldsymbol{A}_\m$ and $s\boldsymbol{c}$,
\begin{align*}
  \boldsymbol{K}^{\m\!} & = K^{a\m}_{\textnormal{\sc T}}\hspace{1pt} T_a
     + K^{a\m}_{\textnormal{\sc Z}}\hspace{1pt} Z_a \,,\\[3pt]
  \boldsymbol{H}^{\m\!} & = H ^{a}_{\textnormal{\sc T}}\hspace{1pt} T_a
     + H^a_{\textnormal{\sc Z}}\hspace{1pt} Z_a \,.
\end{align*}
The path integral~(\ref{ZJ}) becomes
\begin{equation}
   Z\big[ \boldsymbol{J}, \boldsymbol{\bar{\zeta}},  
              \boldsymbol{\zeta}, \boldsymbol{B}; 
              \boldsymbol{K},  \boldsymbol{H}\big] 
     = \int [\rd \boldsymbol{A}]\> [\rd \boldsymbol{c}]\>
             [\rd \boldsymbol{\bar{c}}]\> [\rd \boldsymbol{b}]~
        \exp \hspace{-0.5pt}\Big[ \!-\! \int \!\rd^4\hspace{-0.5pt}x \>
                      {\cal L}_{{\ltimes}}  
        + S_{\rm ext} + S_{\textnormal{KH}} \Big]  \,,
\label{ZJ-total}
\end{equation}
where $S_{\rm KH} $ is given by 
\begin{equation}
    S_{\textnormal{KH}} =  \frac{1}{g^2}  \int \rd^4\hspace{-0.5pt}x\>
      \big[\,  \Om(\boldsymbol{K},s\boldsymbol{A})
         -  \Om(\boldsymbol{H},s\boldsymbol{c})\, \big] 
\label{source-total}
\end{equation}
and $\Om(\boldsymbol{K},\boldsymbol{A})$ and
$\Om(\boldsymbol{H},s\boldsymbol{c})$ read
\begin{align*}
  \Om(\boldsymbol{K},s\boldsymbol{A}) & = \om_{ab} \,
      \big( K^{a\mu}_{ \textnormal{\sc T}}\, 
                 sA^b_{ \textnormal{\sc T}\mu} 
          + K^{a\mu}_{ \textnormal{\sc T}}\, 
                 sA^b_{ \textnormal{\sc Z}\mu} 
          + K^{a\mu}_{ \textnormal{\sc Z}}\, 
                 sA^b_{ \textnormal{\sc T}\mu} \big)\,,     \\[3pt]
\Om(\boldsymbol{H},s\boldsymbol{c}) & = \om_{ab} \,
      \big(H^a_{ \textnormal{\sc T}}\, 
                 sc^b_{ \textnormal{\sc T}} 
          + H^a_{ \textnormal{\sc T}}\, sc^b_{ \textnormal{\sc Z}} 
         +  H^a_{ \textnormal{\sc Z}}\,  sc^b_{ \textnormal{\sc T}}\big)\,.
\end{align*}
The Feynman rules of the theory follow from\, $Z[\cdots]$ in~(\ref{ZJ-total})
and are collected in Figures~1,~2 and~3, where solid lines denote gauge
propagators and dashed lines ghost propagators. The expressions of
$D^{ab}_{\mu\nu}(p)$ and $\Delta^{ab}(p)$ in Figure~1 are
\begin{equation}
      D^{ab}_{\mu\nu}(p)  = g^2\,
       \frac{\dl^{ab}}{p^2}\> \Big[\, \dl_{\m\n} 
           + (\a-1)\,\frac{p_\mu\,p_\nu}{p^2}\, \Big]
\label{gauge-free}
\end{equation}
and
\begin{equation}
      \Delta^{ab}(p) = -\,g^2\>\frac{\dl^{ab}}{p^2}\>.
\label{ghost-free}
\end{equation}
In turn, the gauge vertices $V_{\mu\nu\rho}^{abc}(p,q,k)$ and
$W_{\mu\nu\rho\sigma}^{abcd}$  in Figure~2 read
\begin{equation*}
     V_{\mu\nu\rho}^{abc}(p,q,k) =    
      i f^{abc}\,\big[\,  (q-k)_\mu\,\dl_{\nu\rho} + (k-p)_\nu\,\dl_{\rho\mu} 
          + (p-q)_\rho\,\dl_{\mu\nu} \,\big] 
\end{equation*}
and
\begin{align*}
    W_{\mu\nu\rho\sigma}^{abcd} = - \big[\, &
        f^{abe}\, f^{cde} \, \big( \dl_{\m\rho} \dl_{\n\sig} 
             -  \dl_{\m\sig} \dl_{\n\rho}\big)  
         \nonumber \\
       +  & f^{ace} \, f^{dbe} \, \big( 
         \dl_{\m\sig} \dl_{\rho\n} -  \dl_{\m\n} \dl_{\rho\sig}\big) 
      +  f^{ade} \, f^{bce} \, \big( 
         \dl_{\m\n} \dl_{\sig\rho} -  \dl_{\mu\rho} \dl_{\sig\n}\big)  
      \,\big] \,.  
\end{align*}
We remark that the free propagators do not have {\sc tt} components, and that
their~{\sc tz} component is equal to their {\sc zt} component. This will play
an important part in the analysis below.
\begin{figure}[t]
\begin{center}
\rule{0.85\textwidth}{0.4pt}\\[9pt]
\includegraphics[scale=0.8]{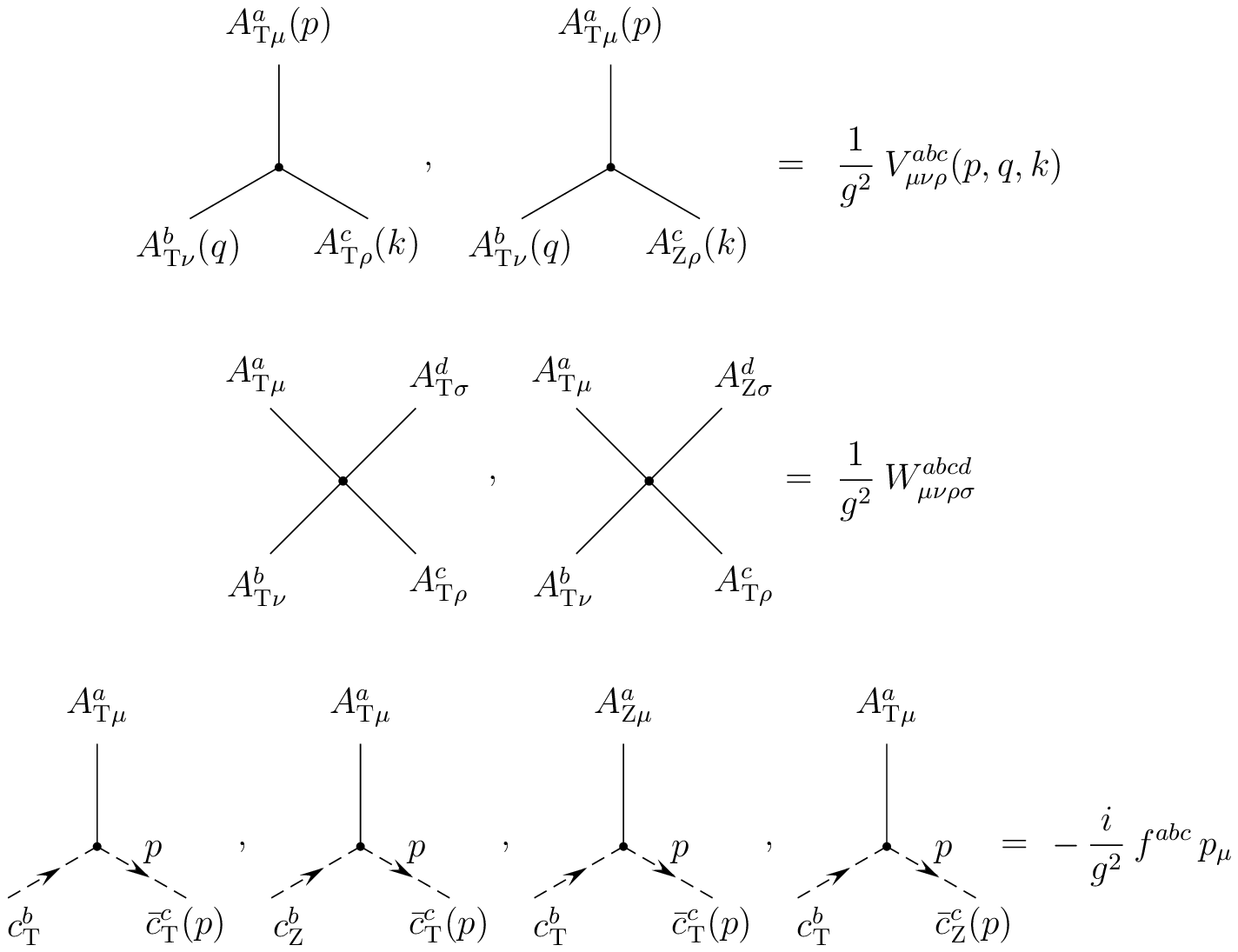} 
\caption{\it Vertices for\, $\textnormal{SU(N)}_{\displaystyle{\ltimes}\!}$
 Yang-Mills theory in Lorenz gauge.}
\rule{0.85\textwidth}{0.4pt}
\vspace{-12pt}
\end{center}
\label{fig:vertices}
\end{figure}
\begin{figure}[h]
\begin{center}
\rule{0.85\textwidth}{0.4pt}\\[12pt]
\includegraphics[scale=0.8]{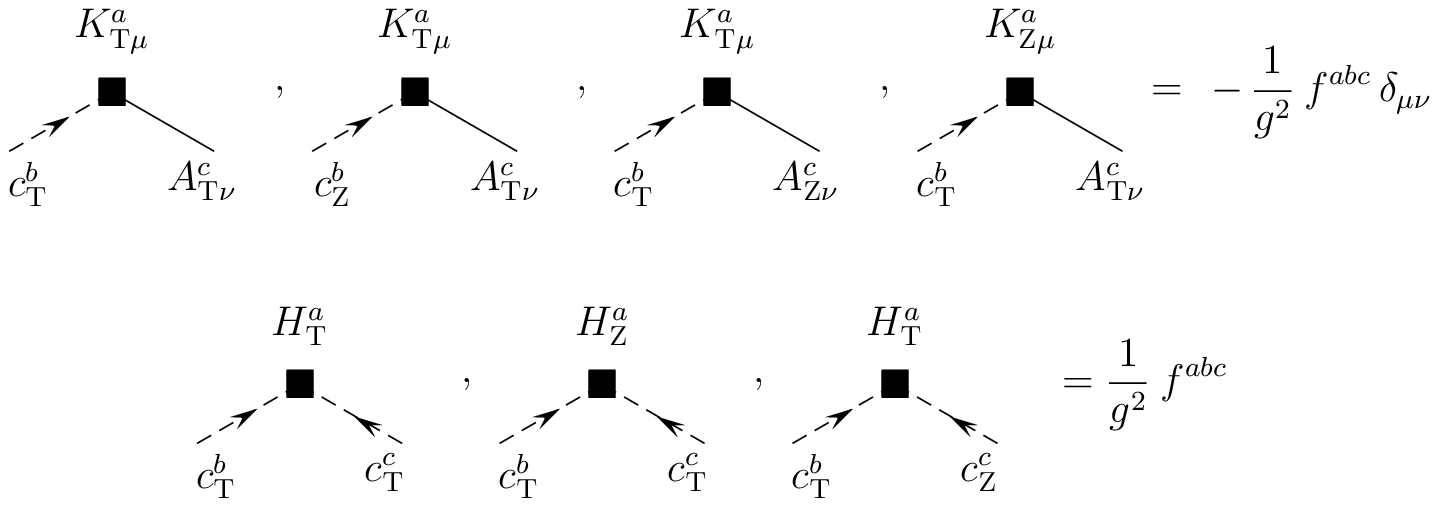}  
\caption{\it External vertices for\, $\textnormal{SU(N)}_{\displaystyle{\ltimes}\!}$
 Yang-Mills theory in Lorenz gauge.}
\rule{0.85\textwidth}{0.4pt}
\vspace{-12pt}
\end{center}
\label{fig:external-vertices}
\end{figure}
\indent Consider for comparison conventional $\textnormal{SU(N)}$ Yang-Mills theory in Lorenz
gauge. Its gauge fixed Lagrangian is recovered from ${\cal L}_\ltimes$ by
setting all the {\sc z} components equal to zero.  To avoid confusion, we
reserve the subscripts {\sc t} and {\sc z} for the field components of the
$\textnormal{SU(N)}_\ltimes$ theory, and use $A^a_\m,\,b^a,\,\bar{c}^a,\,c^a$
and $K^a_\m,\,H^a$ without subscripts for the fields and the nonlinear BRS
sources of the $\textnormal{SU(N)}$ theory. The gauge field and ghost free propagators of the
$\textnormal{SU(N)}$ theory are given by $D^{ab}_{\m\n}(p)$ and
$\Delta^{ab}(p)$ in eqs.~(\ref{gauge-free}) and~(\ref{ghost-free}), which are
equal to the {\sc tz} and {\sc zt} free propagators of the
$\textnormal{SU(N)}_\ltimes$ theory. The Feynman rules for the vertices\,
$A^3,\,A^4,\,\bar{c}Ac,\,KAc$ and $Hcc$ in the $\textnormal{SU(N)}$ theory are as in
Figures~2 and 3.

We now proceed to compute radiative corrections. To regulate whatever UV
divergences may occur, we will use dimensional regularization with
$D=4-2\epsilon$, so from now on all diagrams and Green functions should be
understood as dimensionally regularized. Since dimensional regularization
manifestly preserves BRS invariance, the dimensionally regularized Green
functions will solve the functional identities associated to BRS invariance.

\medskip {\bf One-loop radiative corrections}. The only 1PI one-loop diagrams
that occur in perturbation theory have all their external legs of type {\sc
  t}.  To prove this, note first that the vertices of the theory, see
Figures~2 and~3, have either none or one leg of type~{\sc z}. Assume now that
there is a 1PI one-loop diagram with an external~{\sc z} leg, and call $U_1$
to the vertex to which the leg is attached. All the other legs of $U_1$ will
be of type~{\sc t}. To close a loop, two of these~{\sc t} legs must be
internal. Since there are no {\sc tt} propagators, each internal~{\sc t} leg
must propagate into type {\sc z}. Each one of the resulting {\sc z} legs will
in turn be attached to a different vertex. Call these vertices~$U_2$
and~$U_3$. From~$U_2$ and~$U_3$ only {\sc t} legs will come out. One may go on
and introduce new vertices, but the loop will never close since there are no
{\sc tt} propagators to join two {\sc t} legs. Hence 1PI one-loop diagrams
have all their external legs of type~{\sc t}. The only nonzero 1PI Green
functions at one loop are then $\langle \Psi_{1\textnormal{\sc t}}(p_1) \cdots
\Psi_{n\textnormal{\sc t}}(p_n)\rangle_{\textnormal{SU}(N)_\ltimes}$, where
$\Psi_{i\textnormal{\sc t}}$ stands for any of the fields
$A^a_{\textnormal{\sc T}\mu},\, \bar{c}^a_{\textnormal{\sc T}},\,
c^a_{\textnormal{\sc T}}$ or the sources $K^a_{\textnormal{\sc T}\mu},\,
H^c_{\textnormal{\sc T}}$.
\begin{figure}[h]
\begin{center}
\rule{0.85\textwidth}{0.4pt}\\[9pt]
\includegraphics[scale=0.8]{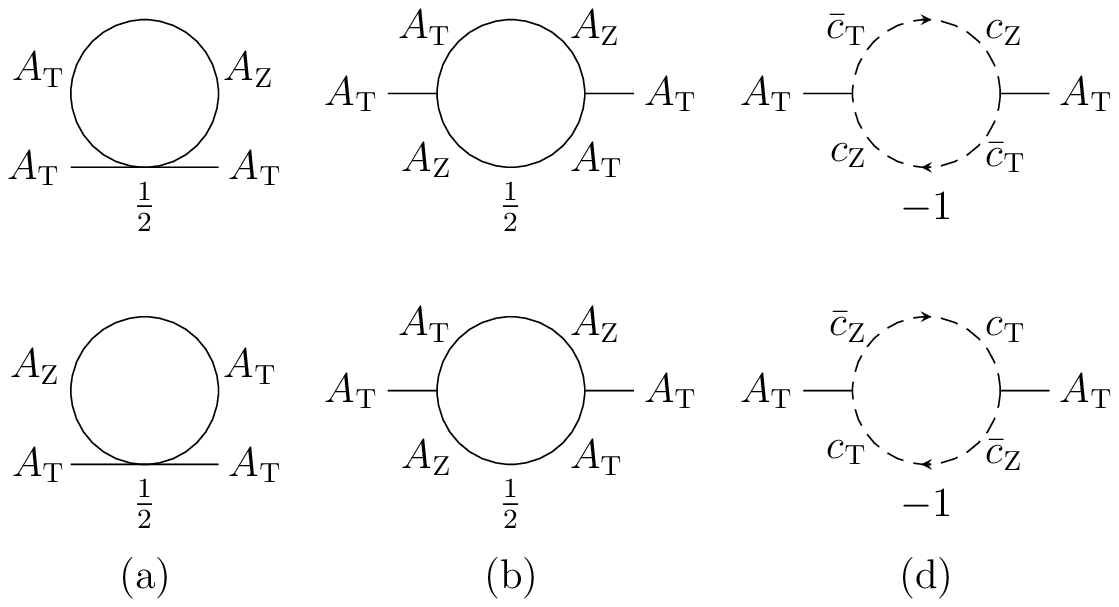} 
\caption{\it One-loop corrections to $\langle A_{\textnormal{\sc
      T}}A_{\textnormal{\sc T}}\rangle_{\textnormal{SU(N)}_\ltimes}$} 
\rule{0.85\textwidth}{0.4pt}
\end{center}
\vspace{-12pt}
\label{fig:two-point-semi}
\end{figure}

Consider for example the two-point 1PI function\, $\langle
A^a_{\textnormal{\sc T}\mu} A^b_{\textnormal{\sc
    T}\nu}\rangle_{\textnormal{SU(N)}_\ltimes}$.  At one loop, it receives
contributions from the diagrams in Figure~4, where the number under each
diagram is the diagram's symmetry factor.  We have drawn in Figure~5 the
one-loop diagrams that contribute to the 1PI function $\langle
A^a_{\mu}A^b_{\nu} \rangle_{\textnormal{SU(N)}}$ of $\textnormal{SU(N)}$
Yang-Mills theory.  Since the propagators and vertices in both sets of
diagrams are the same, we conclude that
\begin{equation}
   \langle A^a_{\textnormal{\sc T}\mu} \hspace{0.5pt}  
       A^b_{\textnormal{\sc T}\nu}\hspace{1pt} 
       \rangle_{\textnormal{SU(N)}_\ltimes}
  = 2\,\langle  A^a_{\mu} A^b_{\nu} \hspace{1pt} 
          \rangle_{\textnormal{SU(N)}} \,.
\label{2-tt}
\end{equation}
Writing only the divergent part as $\eps\to 0$, this gives
 \begin{equation*}
     \langle A^a_{\textnormal{\sc T}\mu}(-p) \hspace{1pt}  
          A^b_{\textnormal{\sc T}\nu}(p) \hspace{1pt} 
          \rangle_{\textnormal{SU}(N)_\ltimes} 
     = \Big(\frac{13}{3}-\a\Big)\,C_\eps\>\Pi_{\mu\nu}^{ab}(p) 
         + O(\epsilon^0) \,,
 \end{equation*}
 where $C_\eps$ is the constant
 \begin{equation}
    C_\eps = -\,\frac{C_2}{16\pi^2\epsilon}\,.
 \label{kappa}
 \end{equation}
\begin{figure}[ht]
\begin{center}
\rule{0.85\textwidth}{0.4pt}\\[9pt]
\includegraphics[scale=0.8]{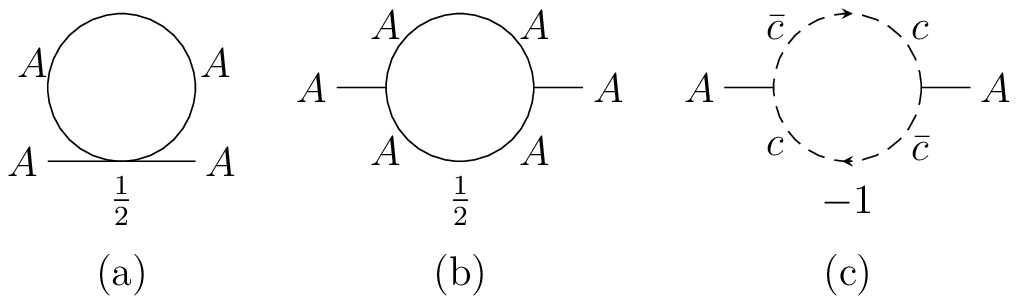} 
\caption{\it One-loop corrections to $\langle A A 
         \rangle_{\textnormal{SU}(N)}$}
\rule{0.85\textwidth}{0.4pt}
\end{center}
\vspace{-12pt}
\label{fig:two-point}
\end{figure}

Eq.~(\ref{2-tt}) can be extended to all 1PI functions as follows. Consider a
1PI one-loop diagram in the $\textnormal{SU(N)}_\ltimes$ theory. Since all its
external legs are of type {\sc t} and there are no {\sc tt} propagators, all
its vertices have one internal leg of type {\sc z}. Label clockwise the
vertices in the loop as $U_1,\ldots,U_n$.  The {\sc z} leg coming out of
vertex $U_1$ must be connected to a internal {\sc t} leg of a neighboring
vertex, say $U_2$, through a {\sc zt} propagator. In turn $U_2$ is connected
to $U_3$ through another {\sc zt} propagator, and so on, until the loop is
closed, with $U_n$ connecting to $U_1$ via a {\sc zt} propagator. For every
such diagram, there is a diagram with the same vertices and the only
difference that 
now the internal {\sc z} leg from $U_1$ connects with an internal {\sc t} leg
in $U_n$ through a {\sc zt} line, rather than with $U_2$. This implies that
$U_n$ connects with $U_{n-1}$ through a {\sc zt} line, and so on until the
loop is closed with a {\sc zt} propagator from $U_2$ with $U_1$.  These two
diagrams give the same contribution to the 1PI function, which in turn is
equal to the contribution of the equivalent 1PI diagram in the
$\textnormal{SU(N)}$ theory. Hence we have
\begin{equation*}
   \langle \Phi_{1,\textnormal{\sc t}}(p_1) \cdots 
        \Phi_{n,\textnormal{\sc t}}(p_n)\rangle_{\textnormal{SU(N)}_\ltimes} 
   = 2\,\langle \Phi_{1}(p_1) \cdots  \Phi_{n}(p_n) 
        \rangle_{\textnormal{SU(N)}} \,.
\end{equation*}
This reduces the calculation of the one-loop 1PI Green functions in the
$\textnormal{SU(N)}_\ltimes$ theory to that in the $\textnormal{SU(N)}$
theory.  The left column In Table~1 collects all the one-loop 1PI Green
functions in $\textnormal{SU(N)}_\ltimes$ Yang-Mills theory that are UV
divergent, whereas the column in the center lists their UV divergent
contributions as computed in dimensional regularization. The column on the
right will be discussed in Section~5.

\medskip
{\bf Vanishing of 1PI radiative corrections beyond one loop}. Any 1PI
\hbox{\emph{n}-loop} diagram can be obtained by joining two external legs in a
1PI \hbox{(\emph{n}-1)-loop} diagram. In our case, since 1PI one-loop diagrams
have all their external legs of type {\sc t} and there are no {\sc tt}
propagators, it is impossible to have two and higher-loop 1PI diagrams. 

We end this section by noting that, again because 1PI Green functions have all
their external legs of type {\sc t} and to these it is only possible to attch
free free {\sc tz} propagators, the only on-shell Green functions that receive
radiative corrections are those having all their external legs of type {\sc
  z}.
\begin{table}[ht]
\begin{center}
\vspace{-3pt}
\renewcommand*{\arraystretch}{1.7}
\begin{tabular}{|l|l|l|} \hline {1PI UV divergent Green function} & {Contribution
    from $\bar{\Gamma}_1^{\eps}$} & {Contribution from}
  $\bar{\Gamma}_1^{\textnormal{ct}}$ \\ \hline
    $\langle \hspace{1pt} A^a_{\textnormal{\sc T}\mu}(-p) \hspace{1pt}   
     A^b_{\textnormal{\sc T}\nu}(p) \hspace{1pt}\rangle $ & 
   $\big( \frac{13}{3} - \a\big) \,C_\eps\,\Pi_{\mu\nu}^{ab}(p)$  & 
   $ -\,\big(\,c_1 + 2c_2\big)\, \Pi^{ab}_{\m\n}(p) $   \\
    $\langle \hspace{1pt} A^a_{\textnormal{\sc T}\mu}(p) \hspace{1pt}
      A^b_{\textnormal{\sc T}\nu}(q) \hspace{1pt}
      A^c_{\textnormal{\sc T}\rho}(k) \hspace{1pt}\rangle$   & 
    $ \big( \frac{17}{6}-\frac{3}{2}\,\a\big) \, 
          C_\eps\, V^{abc}_{\mu\nu\rho}(p,q,k) $ &
    $ -\,(c_1+3c_2)\,V^{abc}_{\mu\nu\rho}(p,q,k)$ \\
    $\langle \hspace{1pt} A^a_{\textnormal{\sc T}\mu}(p) 
      \hspace{1pt} A^b_{\textnormal{\sc T}\nu}(q) \hspace{1pt}
      A^c_{\textnormal{\sc T}\rho}(k) \hspace{1pt}  
      A^d_{\textnormal{\sc T}\sigma}(r) \hspace{1pt}\rangle$ &
   $\big(\frac{4}{3}-2\a\big)\, C_\eps\,W^{abcd}_{\mu\nu\rho\sigma}$ &
   $ -\,\big(c_1 + 4c_2\big)\, 
               W^{abcd}_{\mu\nu\rho\sigma}$ \\
   $\langle \hspace{1pt} \bar{c}^a_{\textnormal{\sc T}}(-p) 
     \hspace{1pt} c^b_{\textnormal{\sc T}}(p)  \hspace{1pt}\rangle$ & 
   $\frac{1}{2}\, (3-\a)\,C_\eps  \dl^{ab} p^2$ & 
   $(c_2-c_3)\, \dl^{ab} p^2$\\
    $\langle \hspace{1pt} \bar{c}^a_{\textnormal{\sc T}}(p) \hspace{1pt}
      A^b_{\textnormal{\sc T}\mu}(q) \hspace{1pt}
      c^c_{\textnormal{\sc T}}(k) \hspace{1pt}\rangle $& 
     $i\a\hspace{1pt} C_\eps\, f^{abc}\,p_\m $ &
     $ic_3 \hspace{1pt} f^{abc}\,p_\m $\\
    $\langle \hspace{1pt} K^a_{\textnormal{\sc T}\m}(-p) 
       \, c^b_{\textnormal{\sc T}}(p) \hspace{1pt}\rangle $ & 
     $\frac{1}{2}\,(3-\a)\,C_\eps\,  \dl^{ab}\,i p_\mu$     & 
    $(c_2-c_3)\,\dl^{ab}\,i p_\mu$\\
    $\langle \hspace{1pt} K^a_{\textnormal{\sc T}\m}(p) \, 
     A^b_{\textnormal{\sc T}\n}(q)\, 
     c^c_{\textnormal{\sc T}}(k) \hspace{1pt}\rangle $ & 
    $-\a\hspace{1pt}C_\eps\,f^{abc}\hspace{1pt} \dl_{\m\n} $  & 
    $-c_3 \hspace{1pt}f^{abc}\hspace{1pt} \dl_{\m\n} $ \\
    $\langle \hspace{1pt} H^a_{\textnormal{\sc T}}(p) \, 
      c^b_{\textnormal{\sc T}}(q)\, 
      c^c_{\textnormal{\sc T}}(k) \hspace{1pt}\rangle $ & 
   $-\,\a \hspace{1pt}C_\eps\,f^{abc}$     &
   $-\,c_3\,f^{abc} $ \\
\hline
\end{tabular}
\caption{\it UV divergent 1PI Green functions in $\textnormal{SU(N)}_\ltimes$
  theory and their counterterms.}
\end{center}
\vspace{-12pt}
\label{table:one-loop-div}
\end{table}

\section{The BRS identity, renormalization and unitarity}

The effective action that generates the 1PI Green functions of
$\textnormal{G}_\ltimes$ Yang-Mills theory is obtained by writing
\begin{equation*}
     Z[\boldsymbol{J}, \boldsymbol{\bar{\zeta}},  
          \boldsymbol{\zeta}, \boldsymbol{B}; \boldsymbol{K},\boldsymbol{H}] 
       = \exp\big(\!-W[\boldsymbol{J}, \boldsymbol{\bar{\zeta}},  
          \boldsymbol{\zeta}, \boldsymbol{B};  \boldsymbol{K},
          \boldsymbol{H}]\,\big)
\end{equation*}
and performing a Legendre transformation on\, $W[\boldsymbol{J},
\boldsymbol{\bar{\zeta}}, \,\boldsymbol{\zeta}, \,\boldsymbol{B};
\boldsymbol{K}, \boldsymbol{H}]$ as follows. Introduce Legendre fields
$\{\boldsymbol{\tilde{A}}, \boldsymbol{\tilde{c}},
\boldsymbol{\tilde{\bar{c}}}, \boldsymbol{\tilde{b}}\}$ for the sources
$\{\boldsymbol{J}, \boldsymbol{\bar{\zeta}}, \boldsymbol{\zeta},
\boldsymbol{B}\}$ through the functional derivatives
\begin{alignat*}{4}
      \tilde{A}^a_{\textnormal{\sc T}\m}(x) &= -\,\om^{ab}\,
         \frac{\dl W}{\dl J^{b\m}_{\textnormal{\sc Z}}(x)}\,, & \qquad
     \tilde{A}^a_{\textnormal{\sc Z}\m}(x) & = \om^{ab}\,\bigg[\,
             \frac{\dl W}{\dl J^{b\m}_{\textnormal{\sc Z}}(x)} 
        - \frac{\dl W}{\dl J^{b\m}_{\textnormal{\sc T}}(x)}\bigg] \,, 
    \\[3pt]
   \tilde{b}^a_{\textnormal{\sc T}}(x) &= -\,\om^{ab}\,
         \frac{\dl W}{\dl B^b_{\textnormal{\sc Z}}(x)}\,, &
     \tilde{b}^a_{\textnormal{\sc Z}}(x) &=  \om^{ab}\,\bigg[\,
         \frac{\dl W}{\dl B^b_{\textnormal{\sc Z}}(x)} 
        - \frac{\dl W}{\dl B^b_{\textnormal{\sc T}}(x)} \bigg] \,,
    \\[3pt]
   \tilde{c}^a_{\textnormal{\sc T}}(x)   &= -\,\om^{ab}\,
    \frac{\dl W}{\dl \bar{\zeta}^b_{\textnormal{\sc Z}}(x)} \,,
     &  \tilde{c}^a_{\textnormal{\sc Z}}(x) & = \om^{ab}\,\bigg[\,
       \frac{\dl W}{\dl \bar{\zeta}^b_{\textnormal{\sc Z}}(x)} 
        - \frac{\dl W}{\dl \bar{\zeta}^b_{\textnormal{\sc T}}(x)} \bigg] \,,
    \\[3pt]
     \tilde{\bar{c}}^a_{\textnormal{\sc T}}(x) &=\om^{ab}\, 
     \frac{\dl W}{\dl \zeta^b_{\textnormal{\sc Z}}(x)} \,, &
   -\,\tilde{\bar{c}}^a_{\textnormal{\sc Z}}(x) &= \om^{ab}\,\bigg[\, 
     \frac{\dl W}{\dl \zeta^b_{\textnormal{\sc Z}}(x)} 
        - \frac{\dl W}{\dl \zeta^b_{\textnormal{\sc T}}(x)}\bigg] \, .
\end{alignat*}
Solve these equations for $\{\boldsymbol{J},
\boldsymbol{\bar{\zeta}}, \boldsymbol{\zeta}, \boldsymbol{B}\}$ in terms of 
fields $\{\boldsymbol{\tilde{A}}, \boldsymbol{\tilde{c}},
\boldsymbol{\tilde{\bar{c}}}, \boldsymbol{\tilde{b}}\}$ and use the solutions
to construct the effective action functional
\begin{align*}
   \Gamma[\boldsymbol{\tilde{A}}, \boldsymbol{\tilde{c}}, 
      \boldsymbol{\tilde{\bar{c}}},  \boldsymbol{\tilde{B}} ;
      \boldsymbol{K}, &\,\boldsymbol{H}] =
   W[\boldsymbol{J}, \boldsymbol{\bar{\zeta}}, \boldsymbol{\zeta}, 
       \boldsymbol{B}; \boldsymbol{K}, \boldsymbol{H}] 
  \nonumber \\[3pt]
   + \int \rd^4\hspace{-1pt}x~& \om_{ab}\,\Big( 
         \, J^a_{\textnormal{\sc T}} \tilde{A}^b_{\textnormal{\sc T}}
      + J^a_{\textnormal{\sc T}} \tilde{A}^b_{\textnormal{\sc Z}}
      + J^a_{\textnormal{\sc Z}} \tilde{A}^b_{\textnormal{\sc T}} 
      + \bar{\zeta}^a_{\textnormal{\sc T}}\,\tilde{c}^b_{\textnormal{\sc T}}
      + \bar{\zeta}^a_{\textnormal{\sc T}}\,\tilde{c}^b_{\textnormal{\sc Z}}
      + \bar{\zeta}^a_{\textnormal{\sc Z}}\,\tilde{c}^b_{\textnormal{\sc
          T}}  
       \nonumber \\[2pt]
      + &\, B^a_{\textnormal{\sc T}}\,\tilde{b}^b_{\textnormal{\sc T}}
      + B^a_{\textnormal{\sc T}}\,\tilde{b}^b_{\textnormal{\sc Z}}
      + B^a_{\textnormal{\sc Z}}\,\tilde{b}^b_{\textnormal{\sc T}}
      + \tilde{\bar{c}}^a_{\textnormal{\sc T}}\,\zeta^b_{\textnormal{\sc T}}
      + \tilde{\bar{c}}^a_{\textnormal{\sc T}}\,\zeta^b_{\textnormal{\sc Z}}
      + \tilde{\bar{c}}^a_{\textnormal{\sc Z}}\,\zeta^b_{\textnormal{\sc
          T}} \Big)\,.
\end{align*}
The very same methods as for Yang-Mills theory with semisimple gauge group show
that~$\Gamma$ has the form
\begin{equation}
   \Gamma = \bar{\Gamma}
    - \int\hspace{-2.5pt} \rd^4\hspace{-0.5pt}x~\om_{ab}\, 
      \Big[ \, \frac{\a}{2}\> \big( 
         \tilde{b}^a_{\textnormal{\sc T}}\,
               \tilde{b}^b_{\textnormal{\sc T}}
        + 2\, \tilde{b}^a_{\textnormal{\sc T}}\,
                \tilde{b}^b_{\textnormal{\sc Z}} \big)      
        + \tilde{b}^a_{\textnormal{\sc T}}\, 
                \pa \tilde{A}^b_{\textnormal{\sc T}} 
        + \tilde{b}^a_{\textnormal{\sc T}}\, 
                \pa \tilde{A}^b_{\textnormal{\sc Z}} 
        + \tilde{b}^a_{\textnormal{\sc Z}}\, 
              \pa \tilde{A}^b_{\textnormal{\sc T}} \Big]\,,
\label{effective}
\end{equation}
where the functional 
\begin{equation*}
    \bar{\Gamma} = \bar{\Gamma}[\boldsymbol{\tilde{A}}_\mu,
      \boldsymbol{\tilde{c}}, \boldsymbol{G}_\mu,
      \boldsymbol{H}] 
\end{equation*}
depends on $\boldsymbol{K}_\m$ and $\boldsymbol{\tilde{\bar{c}}}$ through the
combination
\begin{equation*}
   \boldsymbol{G}_\mu=\boldsymbol{K}_\m 
            + \pa_\mu\boldsymbol{\tilde{\bar{c}}}
\end{equation*}
and satisfies the BRS identity
\begin{equation}
  \int \hspace{-2.5pt} \rd^4\hspace{-1pt} x \; \om^{ab} \,\bigg[\, 
          \frac{\dl \bar{\Gamma}}{\dl \tilde{A}^a_{\textnormal{\sc T}}} \>
         \frac{\dl \bar{\Gamma}}{\dl G^b_{\textnormal{\sc Z}}} 
      + \frac{\dl \bar{\Gamma}}{\dl \tilde{A}^a_{\textnormal{\sc Z}}} \>
         \bigg( \frac{\dl\bar{\Gamma}}{\dl G^b_{\textnormal{\sc T}}}
         - \frac{\dl\bar{\Gamma}}{\dl G^b_{\textnormal{\sc Z}}} 
       \bigg)  
      - \frac{\dl\bar{\Gamma}}{\dl \tilde{c}^a_{\textnormal{\sc T}}} \>
          \frac{\dl\bar{\Gamma}}{\dl H^b_{\textnormal{\sc Z}}} 
      - \frac{\dl\bar{\Gamma}}{\dl \tilde{c}^a_{\textnormal{\sc Z}}} \>
         \bigg( \frac{\dl\bar{\Gamma}}{\dl H^b_{\textnormal{\sc T}}}
         - \frac{\dl\bar{\Gamma}}{\dl H^b_{\textnormal{\sc Z}}}
        \bigg) \bigg] =0\,.
\label{BRS-id}
\end{equation}

The analysis in Section~4 implies that $\bar{\Gamma}$ is the sum 
\begin{equation}
   \bar{\Gamma}  = \bar{\Gamma}_0 + \hbar \,\bar{\Gamma}_1
\label{effectivedr}
\end{equation}
of a tree-level contribution
\begin{align}
     \bar{\Gamma}_0 = \frac{1}{g^2} \int\hspace{-2.5pt} 
        \rd^4\hspace{-1pt}&x ~ \om_{ab} \, \bigg[\,
           \frac{1}{4}\> \tilde{F}^a_{\textnormal{\sc T}} 
                                 \tilde{F}^b_{\textnormal{\sc T}}
       + \frac{1}{2}\> \tilde{F}^a_{\textnormal{\sc T}}
                                   \tilde{F}^b_{\textnormal{\sc Z}}
       - G^a_{\textnormal{\sc T}} \tilde{D}\tilde{c}^b_{\textnormal{\sc T}} 
       -  G^a_{\textnormal{\sc T}}\, \big(
              \tilde{D}\tilde{c}^b_{\textnormal{\sc Z}} + f_{cd}{}^b \,
              \tilde{A}^c_{\textnormal{\sc Z}}\,\tilde{c}^c_{\textnormal{\sc T}}\big)
            \nonumber \\[3pt]
       &  -   G^a_{\textnormal{\sc Z}}\tilde{D}\tilde{c}^b_{\textnormal{\sc T}}
       - \frac{1}{2}\,f_{cd}{}^b \, H^a_{\textnormal{\sc T}}\,    
           \tilde{c}^c_{\textnormal{\sc T}}\hspace{1pt}
           \tilde{c}^d_{\textnormal{\sc T}}
       - f_{cd}{}^b \, H^a_{\textnormal{\sc T}}\, \tilde{c}^c_{\textnormal{\sc T}}
           \hspace{1pt}\tilde{c}^d_{\textnormal{\sc Z}} 
       -   \frac{1}{2}\, f_{cd}{}^b \, H^a_{\textnormal{\sc Z}}\, 
            \tilde{c}^c_{\textnormal{\sc T}}\hspace{1pt}
           \tilde{c}^d_{\textnormal{\sc T}} \,\bigg]
       \label{Gamma0}
\end{align}
and a one-loop contribution $\bar{\Gamma}_1$. The term $\bar{\Gamma}_0$
satisfies the BRS identity~(\ref{BRS-id}).  Substituting
eq.~(\ref{effectivedr}) in eq.~(\ref{BRS-id}), it follows that $\bar{\Gamma}_1
$ must satisfy
\begin{equation*}
    \Delta \bar{\Gamma}_1 =0\,,
\end{equation*}
where $\Delta$ is the Slavnov-Taylor operator 
\begin{align}
    \Delta = \int \hspace{-2.5pt} \rd^4\hspace{-1pt} x \> \om^{ab}
    &\bigg[\,
        \frac{\dl \bar{\Gamma}_0}{\dl \tilde{A}^a_{\textnormal{\sc T}}} \>
        \frac{\dl}{\dl G^b_{\textnormal{\sc Z}}} 
    + \frac{\dl \bar{\Gamma}_0}{\dl G^a_{\textnormal{\sc Z}}} \> 
        \frac{\dl}{\dl \tilde{A}^b_{\textnormal{\sc T}}}  
    + \frac{\dl \bar{\Gamma}_0}{\dl \tilde{A}^a_{\textnormal{\sc Z}}}\>
        \bigg( \frac{\dl}{\dl G^b_{\textnormal{\sc T}}}
               - \frac{\dl}{\dl G^b_{\textnormal{\sc Z}}} \bigg) 
    +  \bigg( \frac{\dl\bar{\Gamma}_0}{\dl G^a_{\textnormal{\sc T}}}
               - \frac{\dl\bar{\Gamma}_0}{\dl G^a_{\textnormal{\sc Z}}} \bigg)\>
       \frac{\dl }{\dl \tilde{A}^b_{\textnormal{\sc Z}}} \nonumber \\[3pt]
    &-  \frac{\dl\bar{\Gamma}_0}{\dl \tilde{c}^a_{\textnormal{\sc T}}} \>
            \frac{\dl}{\dl \tilde{H}^b_{\textnormal{\sc Z}}} 
    - \frac{\dl \bar{\Gamma}_0}{\dl \tilde{H}^a_{\textnormal{\sc Z}}} \>
           \frac{\dl}{\dl \tilde{c}^b_{\textnormal{\sc T}}}   
     - \frac{\dl\bar{\Gamma}_0}{\dl \tilde{c}^a_{\textnormal{\sc Z}}} \>
            \bigg( \frac{\dl}{\dl H^b_{\textnormal{\sc T}}}
                    - \frac{\dl}{\dl H^b_{\textnormal{\sc Z}}} \bigg) 
        - \bigg( \frac{\dl\bar{\Gamma}_0}{\dl H^a_{\textnormal{\sc T}}}
                    - \frac{\dl\bar{\Gamma}_0}{\dl H^a_{\textnormal{\sc Z}}} 
          \bigg) \> \frac{\dl}{\dl \tilde{c}^b_{\textnormal{\sc Z}}} 
          \,\bigg]\,.
\label{Slavnov-Taylor}
\end{align}
The very same arguments as for the semisimple case show that $\Delta$ is
nilpotent, $\Delta^2=0$. The explicit expressions for the action of $\Delta$
on $(\tilde{A}^a_{\textnormal{\sc T}\mu}, \tilde{A}^a_{\textnormal{\sc
    Z}\mu})$, $(\tilde{c}^a_{\textnormal{\sc T}}, \tilde{c}^a_{\textnormal{\sc
    Z}})$ $(G^a_{\textnormal{\sc T}\mu}, G^a_{\textnormal{\sc Z}\mu})$ and
$(H^a_{\textnormal{\sc T}},H^a_{\textnormal{\sc z}})$ are given in the
Appendix.  The operator $\Delta$ is the quantum analog of the BRS operator and
controls gauge invariance for the quantum theory. The only gauge invariant
radiative corrections are those which are cohomologically nontrivial with
respect to $\Delta$. That is, those that cannot be written as $\Delta X$ for
any $X$. Cohomologically trivial corrections are of the form $\Delta\/X$,
originate in gauge fixing and do not contribute to on-shell amplitudes.

We can add to $\bar{\Gamma}_1$ any functional
$\bar{\Gamma}_1^{\textnormal{ct}}$ such that $\Delta
\bar{\Gamma}_1^{\textnormal {ct}\!}=0$. If $\bar{\Gamma}_1^{\textnormal{ct}}$
subtracts the UV divergences in $\bar{\Gamma}_1$, the sum
$\bar{\Gamma}_1^{\,\prime\!}  = \bar{\Gamma}_1 + \bar{\Gamma}_1^{\textnormal
  {ct}}$ \,will be finite and still satisfy $\Delta\bar{\Gamma}_1^{\,\prime\!}
= 0$, thus can be taken as the one-loop contribution to the quantum effective
action. Since the UV divergences in the theory are local, we are interested in
the solution of equation $\Delta \bar{\Gamma}_1^{\textnormal {ct}\!}=0$ over
the space of local integrated functionals of mass dimension four and ghost
number zero\footnote{Both the BRS operator and the Slavnov-Taylor operator
  have mass dimension one and ghost number 1.}. The most general solution over
this space has the form
\begin{equation}
    \bar{\Gamma}_1^{\textnormal {ct}} = c_1  S_{\textnormal{\sc TT}} 
    + \Delta X\,,
\label{ct-invariant} \\[3pt]
\end{equation}
where $c_1$ is an arbitrary real coefficient, $S_{\textnormal{\sc TT}}$ is the
$\textnormal{G}$ Yang-Mills classical action
\begin{equation}
  S_{\textnormal{\sc TT}} = \frac{1}{4} 
        \int \hspace{-2.5pt} \rd^4\hspace{-1pt} x \>\om_{ab}\,
        \tilde{F}^a_{\textnormal{\sc T}} \hspace{1pt} 
        \tilde{F}^b_{\textnormal{\sc T}}
\label{STT}
\end{equation}
and $X$ is any local integrated functional of mass dimension three and ghost
number~$-1$. Note that the cohomologically nontrivial part of the
solution~(\ref{ct-invariant}) does not have a term
\begin{equation}
  S_{\textnormal{\sc tz}} = \frac{1}{2} 
  \int \hspace{-2.5pt} \rd^4\hspace{-1pt} x \>\om_{ab}\,
  \tilde{F}^a_{\textnormal{\sc T}} \hspace{1pt} 
  \tilde{F}^b_{\textnormal{\sc Z}} \,.
\label{STZ}
\end{equation}
This is so since $S_{\textnormal{\sc tz}}$ can be written as $\Delta Y,$ with
$Y$ given by
\begin{equation}
    Y = \int \hspace{-2.5pt} \rd^4\hspace{-1pt} x \> \om_{ab}\,
      \big( G^{a}_{\textnormal{\sc T}} \tilde{A}^b_{\textnormal{\sc Z}} 
            + H^a_{\textnormal{\sc T}} \tilde{c}^b_{\textnormal{\sc
                Z}}\big)\,.
\label{Y-trivial}
\end{equation}
We observe here an important difference between the cohomologies of the BRS
operator~$s$ and the Slavnov-Taylor operator~$\Delta$ over the space of local
integrated functionals of mass dimension four and ghost number zero.  While
$S_{\textnormal{\sc tt}}$ and $S_{\textnormal{\sc tz}}$ are both nontrivial
with respect to $s$, only $S_{\textnormal{\sc tt}}$ is nontrivial with respect
to $\Delta$.

Recall now that the classical action is the sum of the terms
$S_{\textnormal{\sc tz}}$ and $S_{\textnormal{\sc tz}}$. The first one of them
is positive definite and the second one is not. This would seem to point to a
loss of unitarity. This, however, is only apparent since, being
cohomologically trivial with respect to~$\Delta$, $S_{\textnormal{\sc tz}}$
does not carry gauge invariant radiative corrections in the quantum effective
action.

In Section~4 we have used dimensional regularization to compute the one-loop
contribution, call it $\bar{\Gamma}_1^{\textnormal{dreg}}$, to the quantum
effective action for $\textnormal{SU(N)}_\ltimes$. It consists of a divergent
part $\bar{\Gamma}_1^{\eps}$ as $\eps\to\/0$, formed by the terms listed in
the left and center columns in Table 1, and a finite part
$\bar{\Gamma}_1^{\textnormal{fin}}$,
\begin{equation*}
    \bar{\Gamma}_1^{\textnormal{dreg}} = 
         \bar{\Gamma}_1^{\eps} +
              \bar{\Gamma}_1^{\textnormal{fin}}\,.
\end{equation*}
Since dimensional regularization is BRS invariant, both
$\bar{\Gamma}_1^{\eps}$ and $\bar{\Gamma}_1^{\textnormal{fin}}$ satisfy\, $
\Delta\bar{\Gamma}_1^{\eps} = \Delta \bar{\Gamma}_1^{\textnormal{fin}}=0$.  To
remove the UV divergences , we take $\bar{\Gamma}_1^{\textnormal {ct}}$ as in
eq.~(\ref{ct-invariant}), with $X$ given by
\begin{equation}
    X= \int \hspace{-2.5pt} \rd^4\hspace{-1pt} x \> \oṃ_{ab}\,
        \big( c_2 \,G^a_{\textnormal{\sc T}}\hspace{1pt} 
                \tilde{A}^b_{\textnormal{\sc T}}
            + c_3 \,H^a_{\textnormal{\sc T}} \hspace{1pt} 
                \tilde{c}^b_{\textnormal{\sc T}}  \big)
\label{X-trivial}
\end{equation}
and $c_2$ and $c_3$ real coefficients.  The counterterm
$\bar{\Gamma}_1^{\textnormal {ct}}$ then produces the contributions
listed in the right column of Table~1.  We choose $c_1,\,c_2$ and~$c_3$ so
that $\bar{\Gamma}_1^{\eps}+\bar{\Gamma}_1^{\textnormal {ct}}=0$. This
corresponds to a minimal subtraction scheme and defines a finite renormalized
effective action
\begin{equation*}
   \bar{\Gamma} =  \lim_{\eps\to\/0} \big( 
             \bar{\Gamma}_0 + \bar{\Gamma}_1^{\textnormal{fin}}\big)\,.
\label{renormalized}
\end{equation*}
Using the results in Table~1, we have
\begin{equation}
  c_1= \frac{22}{3}\,C_\eps  \, \quad c_2= -\,\frac{\a+3}{2}\,C_\eps\,,
  \quad c_3= -\a\hspace{1.5pt}C_\eps \,,
\label{coefficients}
\end{equation}
with $C_\eps$ as in eq.(\ref{kappa}). These values for $c_1,\,c_2$ and $c_2$
are twice those for $\textnormal{SU(N)}$ Yang-Mills theory. This implies in
particular that the first coefficient of the beta function for
$\textnormal{SU(N)}_\ltimes$ is $-22/3$, rather than the usual $-11/3$.

\medskip{\bf Multiplicative renormalization}. The subtraction performed by the
counterterm $\bar{\Gamma}_1^{\rm ct}$ in eqs.~(\ref{ct-invariant})
and~(\ref{X-trivial}) is equivalent to multiplicative renormalization.  To see
this, recall that in multiplicative renormalization, the fields and the
coupling constant in the tree-level action $\bar{\Gamma}_0$ in
eq.~(\ref{Gamma0}) are regarded as bare fields $\{\boldsymbol{A}_{0\mu},
\boldsymbol{c}_0, \boldsymbol{G}_{0\mu}, \boldsymbol{H}_0\}$ and bare coupling
constant $g_0$. Renormalized quantities are then introduced through the
equations
\begin{alignat*}{10}
   A^{~a}_{0\textnormal{\sc T}\mu} & = 
          Z^{\textnormal{\sc T}}_A\,\tilde{A}^a_{\textnormal{\sc T}\mu}\,,
  &\qquad & 
 c^{~a}_{0\textnormal{\sc T}} & = 
          Z^{\textnormal{\sc T}}_c\,\tilde{c}^a_{\textnormal{\sc T}}\,,
  &\qquad &
  G^{~a}_{0\textnormal{\sc T}} & = 
          Z^{\textnormal{\sc T}}_G\, G^a_{\textnormal{\sc T}}\,,
  &\qquad & 
  H^a_{0\textnormal{\sc T}} & = 
          Z^{\textnormal{\sc T}}_G\, H^a_{\textnormal{\sc T}}\,,\\[3pt]
  A^{~a}_{0\textnormal{\sc Z}\mu} & = 
          Z^{\textnormal{\sc Z}}_A\,\tilde{A}^a_{\textnormal{\sc Z}\mu}\,,
  &\qquad & 
  c^{~a}_{0\textnormal{\sc Z}} & = 
          Z^{\textnormal{\sc Z}}_c\,\tilde{c}^a_{\textnormal{\sc Z}}\,,
  &\qquad &
  G^{~a}_{0\textnormal{\sc Z}} & = 
          Z^{\textnormal{\sc Z}}_G\,G^a_{\textnormal{\sc Z}\mu}\,,
  &\qquad & 
  H^{~a}_{0\textnormal{\sc Z}} & = 
          Z^{\textnormal{\sc Z}}_G\,H^a_{\textnormal{\sc Z}}
\end{alignat*}
and 
\begin{equation*}
  g_0 = Z_g g\,.
\end{equation*}
Writing every renormalization constant as $Z=1+\dl\/Z$, with $\dl\/Z$
first-order in perturbation theory, the action
$\bar{\Gamma}_0[\boldsymbol{\Psi}_0,g_0]$ is recast as
\begin{equation*}
  \bar{\Gamma}_0[ \boldsymbol{\tilde{A}}_0, \boldsymbol{\tilde{c}}_0, 
                                      \boldsymbol{G}_0, \boldsymbol{H}_0,g_0] =
  \bar{\Gamma}_0[ \boldsymbol{\tilde{A}}, \boldsymbol{\tilde{c}}, 
                                      \boldsymbol{G}, \boldsymbol{H},g] +
  \dl\/\bar{\Gamma}_0[ \boldsymbol{\tilde{A}}, \boldsymbol{\tilde{c}}, 
                                      \boldsymbol{G}, \boldsymbol{H},g ]\,,
\end{equation*}
where the counterterm $\dl\/\bar{\Gamma}_0[\cdots]$ collects all contributions
of order one,
\begin{equation*}
  \dl\bar{\Gamma}_0 [ \boldsymbol{\tilde{A}}, \boldsymbol{\tilde{c}}, 
        \boldsymbol{G}, \boldsymbol{H},g] 
   = \frac{1}{g^2} \int \hspace{-2.5pt} \rd^4\hspace{-1pt} x \;
       \om_{ab}\,\Big[\, 
         \frac{1}{2}\>(2\,\dl Z_A^{\textnormal{T}} -2\,\dl Z_g) \,
         \tilde{A}^{a\m}_{\textnormal{\sc T}}\, \big(\dl_{\m\n}\pa^2-\pa_\m\pa_\n\big)  
         \tilde{A}^{b\n}_{\textnormal{\sc T}}   +\cdots \Big]\,. 
\end{equation*}
The requirement that $\dl\bar{\Gamma}_0 [\boldsymbol{\tilde{\Psi}},g]$ must
cancel the UV divergences fixes
\begin{equation*}
    \begin{array}{rlcrl}
       \dl\/Z_g&\hspace{-7pt}= -\,{\displaystyle \frac{g^2c_1}{2}}\,,& &
             \dl\/Z_c^{\textnormal{\sc T}} + \dl\/Z_G^{\textnormal{\sc T}}
                 & \hspace{-7pt}= - g^2(c_1+c_2+c_3)\,, \\[6pt]
      \dl\/Z_A^{\textnormal{\sc T}}& \hspace{-7pt}= g^2c_2\,, &  &
         \dl\/Z_H^{\textnormal{\sc T}} + 2\,\dl\/Z_c^{\textnormal{\sc T}} 
        & \hspace{-7pt}= g^2(c_3-c_1)\,,
   \end{array}
\end{equation*}
where $c_1, c_2$ and $c_3$ are the coefficients in the right column in
Table~1. Since there are no one-loop 1PI diagrams with {\sc z} external legs,
there are no conditions for $\dl\/Z^{\textnormal{\sc Z}}_\Psi$.  The question
then arises as to what is the meaning of the {\sc z} terms in
$\dl\/\bar{\Gamma}_0[\boldsymbol{\tilde{\Psi}},g]$. Our analysis above
indicates that they should be cohomologically trivial with respect to the
Slavnov-Taylor operator. And this is indeed the case. A long but
straightforward calculation shows that, for
\begin{equation*}
     \dl\/Z_A^{\textnormal{\sc Z}} = \dl\/Z_A^{\textnormal{\sc T}}\,,\quad
     \dl\/Z_c^{\textnormal{\sc Z}} = \dl\/Z_c^{\textnormal{\sc T}}\,,\quad
     \dl\/Z_G^{\textnormal{\sc Z}} = \dl\/Z_G^{\textnormal{\sc T}}\,,\quad
     \dl\/Z_H^{\textnormal{\sc Z}} = \dl\/Z_H^{\textnormal{\sc T}}\,,
\end{equation*}
the functional\, $\dl\bar{\Gamma}_0 [ \boldsymbol{\tilde{A}},
\boldsymbol{\tilde{c}}, \boldsymbol{G}, \boldsymbol{H},g]$ \,can be written as
\begin{equation}
    \dl\bar{\Gamma}_0 [ \boldsymbol{\tilde{A}}, \boldsymbol{\tilde{c}}, 
        \boldsymbol{G}, \boldsymbol{H},g]
         = c_1S_{\textnormal{\sc TT}} +\Delta (X+U)\,,
\label{ct-2}
\end{equation}
where $X$ is given by eq.~(\ref{X-trivial}) and  $U$ has the form
\begin{equation*}
    U = \int \hspace{-2.5pt} \rd^4\hspace{-1pt} x \>\om_{ab}\,\Big[
      (c_1+c_2)\,  G^a_{\textnormal{\sc T}} 
          \tilde{A}^b_{\textnormal{\sc Z}}-
     (c_1+2c_2+c_3)\,  H^a_{\textnormal{\sc T}} 
          \tilde{c}^b_{\textnormal{\sc Z}}  +
      c_2\,  G^a_{\textnormal{\sc Z}} 
         \tilde{A}^b_{\textnormal{\sc T}} -
      c_3\,  H^a_{\textnormal{\sc Z}} 
         \tilde{c}^b_{\textnormal{\sc T}} \Big]\,.
\end{equation*}
The difference between the counterterms $\dl\bar{\Gamma}_0$ above and
$\bar{\Gamma}_1^{\rm ct}$ in eqs.~(\ref{ct-invariant}) and~(\ref{X-trivial})
is $\Delta\/U$, which is cohomologically trivial.

All in all, the only gauge invariant radiative corrections are those in $c_1$,
which account for a renormalization of the coupling constant. This introduces
a renormalization scale in the quantum effective action and the quantum theory
is asymptotically free.

\section{Discussion}

The pattern observed for the gauge invariant degrees of freedom in the quantum
theory resembles very much that for the self-antiself dual instantons of the
classical theory~\cite{FRR}.  In the classical case, the number of collective
coordinates of the $\textnormal{G}_\ltimes$ instantons is twice that of the
embedded $\textnormal{G}$ instantons, yet\, $\om_{ab}F^a_{\textnormal{\sc
    T}\mu\nu}F^{b\mu\nu}_{\textnormal{\sc Z}}$ \,does not contribute to the
instanton number. Now the gauge invariant radiative corrections are doubled
and $\om_{ab}F^a_{\textnormal{\sc T}\mu\nu}F^{b\mu\nu}_{\textnormal{\sc Z}}$
is cohomologically trivial with respect to the Slavnov-Taylor operator.

Our discussion may have some implications for Yang-Mills theories with more
general nonreductive real metric Lie algebras. There is a structure
theorem~\cite{Medina-Revoy} that states that all real metric Lie algebras are
direct products of Abelian algebras, simple real Lie algebras and double
extensions $\mathfrak{d}(\mathfrak{h},\mathfrak{g})$ of a real metric Lie
algebra $\mathfrak{h}$ by an algebra $\mathfrak{g}$\footnote{The theorem goes
  further and specifies the nature of $\mathfrak{g}$ in the double
  extension.}.  The double extension $\mathfrak{d}(\mathfrak{h},\mathfrak{g})$
is obtained~\cite{Medina-Revoy,FO-Stanciu-double} by forming the classical
double $\mathfrak{g}_\ltimes$ and then by acting with $\mathfrak{g}$ on
$\mathfrak{h}$ via antisymmetric derivations. Incidentally we mention that the
classical double $\mathfrak{g}_\ltimes$ can be viewed as the double extension
of the trivial algebra by $\mathfrak{g}$.

According to the theorem, since $\mathfrak{h}$ must be metric, three
possibilities must be considered for $\mathfrak{h}$ in forming double
extensions $\mathfrak{d}(\mathfrak{h},\mathfrak{g})$. The first one is that
$\mathfrak{h}$ is a simple real Lie algebra. In this
case~\cite{FO-Stanciu-double}, the algebra of antisymmetric derivations of
$\mathfrak{h}$ is $\mathfrak{h}$ itself and the double extension
$\mathfrak{d}(\mathfrak{h},\mathfrak{h})$ is isomorphic to the direct product
$\mathfrak{h}\times \mathfrak{h}_\ltimes$. The resulting Yang-Mills theory
then separates into two Yang-Mills theories, not interacting with each other,
one with gauge group $\textnormal{H}$ and one with group
$\textnormal{H}_\ltimes$. The second possibility is that $\mathfrak{h}$ is
Abelian, of dimension $m$. Being Abelian, any nondegenerate, symmetric
bilinear form on $\mathfrak{h}$ is a metric, and can always be brought to a
diagonal form with all the entries in the diagonal equal to $+1$ and $-1$. If
the number of occurrences of $+1$ is $p$, and the number of occurrence of $-1$
is $q$, the algebra of antisymmetric derivations of $\mathfrak{h}$ is any
subalgebra of $\as\ao(p,q)$~\cite{FO-Stanciu-double}. Many of the
nonsemisimple WZW models considered in the literature~\cite{Nappi-Witten,
  Sfetsos-1, Sfetsos-2, Sfetsos-3, Mohammedi,FO-Stanciu-nonreductive} and
their four-dimensional Yang-Mills analogs~\cite{Tseytlin} fall into this
class. In this instance unitarity remains an open problem. We think that a
thorough analysis of the corresponding Slavnov-Taylor operator should shed
some light on the problem. The third possibility for $\mathfrak{h}$ is that it
is a double extension, which takes us back to the starting point.

\numberwithin{equation}{section}
\renewcommand{\theequation}{{A}.\arabic{equation}}
\appendix
\section*{Appendix}

The action of the Slavnov-Taylor operator~(\ref{Slavnov-Taylor}) on the fields
$(\tilde{A}^a_{\textnormal{\sc T}\mu}, \tilde{A}^a_{\textnormal{\sc Z}\mu})$
and $(\tilde{c}^a_{\textnormal{\sc T}}, \tilde{c}^a_{\textnormal{\sc Z}})$ is
given by
\begin{align*}
  \Delta\tilde{A}^a_{\textnormal{\sc T}\mu} 
         & = -\,\tilde{D}_\mu\/\tilde{c}^a_{\textnormal{\sc T}}\, ,\\[3pt]
  \Delta\tilde{A}^a_{\textnormal{\sc Z}\mu} & = -\,\big(
       \tilde{D}_\mu \tilde{c}^a_{\textnormal{\sc T}} 
         + f_{bc}{}^{a}\,\tilde{A}^b_{\textnormal{\sc Z}\mu}\, 
              \tilde{c}^c_{\textnormal{\sc  T}}\big)\, ,\\[3pt] 
  \Delta\tilde{c}^a_{\textnormal{\sc T}} & =\frac{1}{2}\> f_{bc}{}^{a}\,
        \tilde{c}^b_{\textnormal{\sc  T}} \,
               \tilde{c}^c_{\textnormal{\sc  T}} \,,\\[3pt]
  \Delta\tilde{c}^a_{\textnormal{\sc Z}} &  =  f_{bc}{}^{a}\,
        \tilde{c}^b_{\textnormal{\sc  T}} \, 
             \tilde{c}^c_{\textnormal{\sc  Z}} \,,
\end{align*}
Modulo an irrelevant overall sign, these expressions generalize the classical BRS
operator $s$ in eqs.~(\ref{BRS-T}) and (\ref{BRS-Z}). The action on
$(G^a_{\textnormal{\sc T}\mu}, G^a_{\textnormal{\sc Z}\mu})$ and
$(H^a_{\textnormal{\sc T}},H^a_{\textnormal{\sc z}})$ is in turn
\begin{align*}
  \Delta\/G^a_{\textnormal{\sc T}\mu} & = -\,\big( 
       \tilde{D}^\rho\/\tilde{F}^a_{\textnormal{\sc T}\rho\mu} 
           + f^{abc}\,G^b_{\textnormal{\sc T}\mu} \,  
                \tilde{c}^c_{\textnormal{\sc  T}}\big)\,,\\[3pt]
  \Delta\/G^a_{\textnormal{\sc Z}\mu} & = -\,\big(
         \tilde{D}^\rho\/\tilde{F}^a_{\textnormal{\sc Z}\rho\mu} 
       + f_{bc}{}^{a}\,\tilde{A}^{b\rho}_{\textnormal{\sc Z}}\, 
                \tilde{F}^c_{\textnormal{\sc  T}\rho\mu}\big) 
      + f_{bc}{}^{a}\, \big( G^b_{\textnormal{\sc Z}\mu} \, 
                                \tilde{c}^c_{\textnormal{\sc  T}}  
                            + G^b_{\textnormal{\sc T}\mu} \, 
                                \tilde{c}^c_{\textnormal{\sc  Z}} \big) \, ,\\[3pt] 
  \Delta\/H^a_{\textnormal{\sc T}} & = 
        \tilde{D}_\mu  G^{a\mu}_{\textnormal{\sc T}} 
      - f_{bc}{}^{a}\,H^b_{\textnormal{\sc T}}\, 
         \tilde{c}^c_{\textnormal{\sc  T}} \,,\\[3pt]
  \Delta\/H^a_{\textnormal{\sc Z}} &  =  
        \tilde{D}_\mu  G^{a\mu}_{\textnormal{\sc Z}} 
       - f_{bc}{}^{a}\, G^{b\mu}_{\textnormal{\sc T}\mu}
         \tilde{A}^{c\mu}_{\textnormal{\sc Z}}
      - f_{bc}{}^{a}\, \big( H^b_{\textnormal{\sc Z}}\, 
                             \tilde{c}^c_{\textnormal{\sc  T}} 
                          + H^b_{\textnormal{\sc T}}\, 
                             \tilde{c}^c_{\textnormal{\sc  Z}} \big)\,. 
\end{align*}
With this, it is matter of algebra to check that (i) the action of $\Delta$ on
$\bar{\Gamma}_1^{\rm ct}$ in eqs.~(\ref{ct-invariant}) and~(\ref{X-trivial})
produces the terms in the right column of Table~1, (ii) that
$S_{\textnormal{\sc TZ}}$ in eq.~(\ref{STZ}) equals $\Delta\/Y$, with $Y$
given by eq.~(\ref{Y-trivial}), and (iii) that $\dl\bar{\Gamma}_0 [
\boldsymbol{\tilde{A}}, \boldsymbol{\tilde{c}}, \boldsymbol{G},
\boldsymbol{H},g]$ can be written as in eq.~(\ref{ct-2}).

\section*{Acknowledgment}

This work was partially funded by the Spanish Ministry of Economy and
Competitiveness  through grant FPA2014-54154-P and by the European Union Cost
Program through grant MP 1405.

\end{document}